\documentclass[journal]{IEEEtran}

\usepackage{pdfpages} 
\usepackage{mathtools}
\usepackage{pgfplots}
\usepackage{pgfplotstable}
\usepackage{pifont}

\usepackage{subfig}

\usepackage{xcolor}
\usepackage{soul}

\usepackage{comment}

\usepackage{graphicx}
\usepackage{multirow}
\usepackage{siunitx}
\usepackage{multirow}

\newcommand{\subparagraph}{}
\usepackage{authblk}
\usepackage{amssymb,amsmath}
\usepackage[english]{babel}
\usepackage{color,soul}
\usepackage{amsthm}
\makeatletter

\let\proof\@undefined
\let\endproof\@undefined
\makeatother

\newtheorem{mydef}{Definition}

\newcolumntype{P}[1]{>{\centering\arraybackslash}p{#1}}
\newcolumntype{M}[1]{>{\centering\arraybackslash}m{#1}}

\usepackage{cite}

\makeatletter
\def\hlinewd#1{%
\noalign{\ifnum0=`}\fi\hrule \@height #1 %
\futurelet\reserved@a\@xhline}
\makeatother
\usepackage[tworuled,noline,linesnumbered]{algorithm2e}
\SetNlSkip{0.9em}
\SetNlSty{textbf}{L}{}

\SetAlFnt{\footnotesize}
\SetAlCapFnt{\footnotesize}
\SetAlCapNameFnt{\footnotesize}

\usepackage{amssymb,amsmath}

\usepackage{amsmath}

\usepackage[ampersand]{easylist}
\ListProperties(Hang=true,Hide=1000,
Margin1=0.42em,Style1*=\textbullet\hskip .5em,
Margin2=3.7em,Style2*=--\hskip .5em,
Margin3=5.9em,Style3*=$\ast$\hskip .5em,
Margin4=7.8em,Style4*=$\cdot$\hskip .5em)

\usepackage[compact]{titlesec}
\titlespacing{\section}{1pt}{2pt}{2pt}
\titlespacing{\subsection}{1pt}{2pt}{2pt}
\titlespacing{\subsubsection}{1pt}{2pt}{2pt}

\newcommand{\figurefontsize}{\footnotesize}

\graphicspath{{./Fig/}}
\newcommand{\tum}{Technical University of Munich (TUM)}

\def\figname{Fig.}

\newcommand{\reviewhl}[1]{\textcolor{black}{#1}}
\newcommand{\hlreview}[1]{\textcolor{black}{#1}}

\begin{document}

\newcommand{\titlenym}{EffiTest2}
\newcommand{\papertitle}{TimingCamouflage+: Netlist Security Enhancement with Unconventional Timing}

\title{\papertitle}

\author{
Grace Li Zhang, Bing Li, Meng Li, Bei Yu, David Z. Pan, Michaela Brunner, Georg Sigl and Ulf Schlichtmann \vspace{-2em}
 \thanks{ %
This research was partly funded by the Fraunhofer High Performance Center Secure Connected Systems Munich. 
        }
\thanks{A preliminary version of this paper was published in
the Proceedings of the Design, Automation, and Test in Europe (DATE)
conference, 2018\cite{ZLYPS18}.}
        \thanks{Grace Li Zhang, Bing Li, and Ulf Schlichtmann are with the Chair of
Electronic Design Automation,  
        Technical University of Munich   
 (e-mail: grace-li.zhang@tum.de; b.li@tum.de;
ulf.schlichtmann@tum.de).}
        \thanks{Bei Yu is with the Department of Computer Science and Engineering, The Chinese University of Hong Kong
 (e-mail: byu@cse.cuhk.edu.hk).}
         \thanks{Meng Li and David Z. Pan are with the
 Department of Electrical and Computer Engineering,  University of Texas at Austin 
 (e-mail: meng\_li@utexas.edu; dpan@ece.utexas.edu).}
           \thanks{Michaela Brunner and Georg Sigl are with the Chair of 
Security in Information Technology, Technical University of Munich 
(e-mail: michaela.brunner@tum.de; sigl@tum.de).}

}

\maketitle
\markboth{IEEE TRANSACTIONS ON COMPUTER-AIDED DESIGN OF INTEGRATED CIRCUITS AND SYSTEMS}
{Zhang \MakeLowercase{\textit{et al.}}: \papertitle}

\IEEEpeerreviewmaketitle

\begin{abstract}
With recent advances in reverse engineering, 
attackers can reconstruct a netlist to counterfeit chips
by opening the die and scanning all layers of authentic chips.
This relatively easy counterfeiting is made possible
by the use of the standard
simple clocking scheme, where all combinational blocks 
function within one clock period,
so that a netlist of combinational logic gates and flip-flops 
is sufficient to duplicate a design. 
In this paper, we propose 
to invalidate the
assumption that a netlist completely represents the function of a
circuit with unconventional timing.  
With the introduced wave-pipelining paths,  
attackers have to capture 
gate and interconnect delays during reverse engineering, or 
to test a huge number of combinational paths to identify the wave-pipelining paths. 
To hinder the test-based attack,
we construct false paths with wave-pipelining to increase
the counterfeiting challenge.
Experimental results confirm that wave-pipelining true paths and false paths   
can be constructed in benchmark circuits 
successfully with only a negligible cost, thus thwarting  %
the potential attack techniques. %

\end{abstract}

\section{Introduction}

A major IC (Integrated Circuit) counterfeiting threat is the illegal production of chips 
by a third party with a netlist 
reverse engineered from authentic chips. %
In reverse
engineering,
authentic chips are delayered and imaged to
identify logic gates, flip-flops, and their connections to reconstruct a
netlist.
Because the recognized netlist carries all necessary design information,
this reverse engineering flow allows counterfeiters to 
reproduce authentic chips with much freedom.

Several techniques have been proposed to thwart reverse engineering attacks on
authentic chips.
The first method is IC camouflaging which tries to
prevent the netlist from being recognized easily \cite{Arunk17}.
In \cite{BeckerRPB14} transistors are manipulated with a
stealthy doping technique during manufacturing so that they function
differently than they appear. 
In \cite{Lap07}, the layouts of different cells are designed to be identical,  
leading
to difficulty in interpreting the functionality of a netlist.
The work in
\cite{MalikBPB15,Rajendran2013,RajendranSK13} mixes real and dummy contacts 
to camouflage standard cells so that they cannot be recognized by reverse 
engineering. The method in \cite{LeeT15} explores netlist obfuscation 
by iterative logic fanin cone analysis at circuit level. In addition, the method
in \cite{LiSMZYJP16} introduces a quantitative security criterion and proposes
camouflaging techniques with a low-overhead cell library and
an AND-tree structure to strengthen netlist security. 
The functionality of a given design is perturbed in \cite{Muham16} 
by applying a simple transformation and   
a separate camouflaged block is used to 
recover the functionality. %
Moreover, the method in \cite{KavehP17} creates logic loops by adding dummy wires 
and gates to obfuscate the circuit topology. 
However, nearly all these camouflaging methods can be deobfuscated 
by attacks based on 
 Boolean Satisfiability (SAT) \cite{Kaveh2017,Yuanqi2017,Abhi16}.

The second method to thwart reverse engineering 
is logic locking, which inserts additional logic gates, e.g.,
XOR/XNOR in \cite{Rajendran2012,RoyKM08}, 
AND/OR in \cite{DupuisBNFR14}, MUX in \cite{PlazaM15} and look-up tables (LUTs) in \cite{Alex2010}, 
into 
the netlist. These components can only activate the correct 
function of the circuit with a given key. 
This method is expanded in \cite{YA17} to incorporate delay information into the locking mechanism. 
Furthermore, logic locking can 
be performed at sequential level to prevent the circuit 
from entering working states without a valid key \cite{ChakrabortyB09}. 
Recently, logic locking has been applied to protect 
parametric behavior of circuits, e.g., pipelined processors\cite{Monir2018}, 
GPUs\cite{GPU18} and analog circuits\cite{Nith2018}. 
The method in \cite{Monir2018} adds meaningless clock cycles to camouflage a design, where 
only correct keys allow a high timing performance. 
However, various attack methods, e.g., SAT attack\cite{Pra2015}, 
removal attack\cite{MMM2017} and bypass attack\cite{Xiaolin2017} 
can potentially recognize the correct 
keys. 

Beyond the techniques described above, 
other methods, e.g., watermarking\cite{Dkiro03}, metering\cite{Fari2001} and split 
manufacturing\cite{xiaoxiao2017,Jeya2013}, can also 
be applied against counterfeiting. 
But the purpose of these methods are primarily to prevent overproduction 
and they need to be adapted properly to counter reverse engineering.

The existing methods of circuit protection either make the netlist more
difficult to be recognized, or make the correct behavior of the circuit
dependent on additional input information even after the netlist is
recognized.
In this paper, we propose a new perspective to 
counter counterfeiting based on reverse engineering. 
Our contributions are as follows. 
\begin{easylist}
&  A new dimension of IC camouflage is proposed to secure circuit netlists 
by integrating unconventional timing 
information. 
A camouflaged netlist thus only works correctly with a 
given set of timing information, which, however, 
is difficult to be recognized exactly by reverse engineering. 
 
& To integrate unconventional timing into a netlist, 
wave-pipelining paths are constructed 
in some parts of a circuit. 
To prevent the exposure of these paths resulting from clustering, 
the retiming technique is deployed to 
spread them across a circuit by blocking the paths not related to wave-pipelining construction.

\begin{figure}[t]
{\figurefontsize
\centering
\begingroup%
  \makeatletter%
  \providecommand\color[2][]{%
    \errmessage{(Inkscape) Color is used for the text in Inkscape, but the package 'color.sty' is not loaded}%
    \renewcommand\color[2][]{}%
  }%
  \providecommand\transparent[1]{%
    \errmessage{(Inkscape) Transparency is used (non-zero) for the text in Inkscape, but the package 'transparent.sty' is not loaded}%
    \renewcommand\transparent[1]{}%
  }%
  \providecommand\rotatebox[2]{#2}%
  \ifx\svgwidth\undefined%
    \setlength{\unitlength}{219.31569459bp}%
    \ifx\svgscale\undefined%
      \relax%
    \else%
      \setlength{\unitlength}{\unitlength * \real{\svgscale}}%
    \fi%
  \else%
    \setlength{\unitlength}{\svgwidth}%
  \fi%
  \global\let\svgwidth\undefined%
  \global\let\svgscale\undefined%
  \makeatother%
  \begin{picture}(1,0.83527493)%
    \put(0.0059012,0.38728013){\color[rgb]{0,0,0}\makebox(0,0)[lb]{\smash{}}}%
    \put(0,0){\includegraphics[width=\unitlength,page=1]{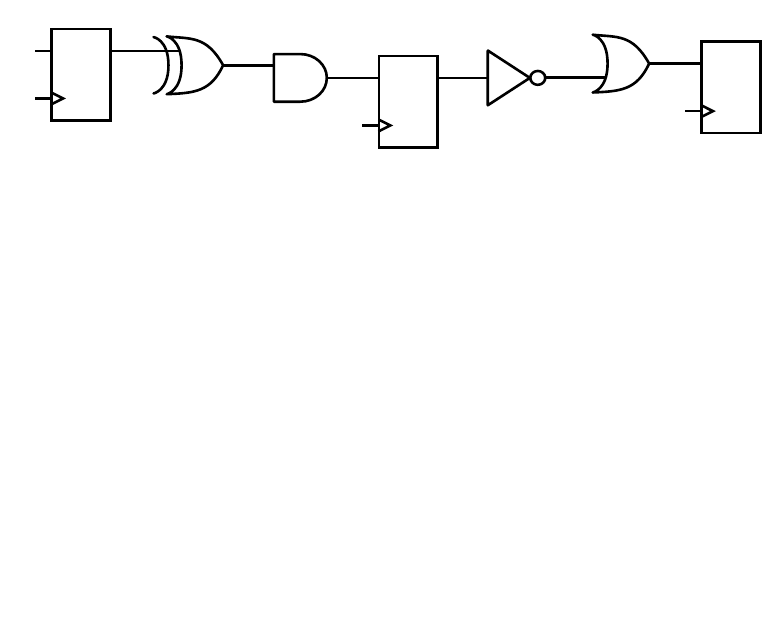}}%
    \put(0.10871432,0.72323846){\color[rgb]{0,0,0}\makebox(0,0)[b]{\smash{F1}}}%
    \put(0.53539391,0.68774384){\color[rgb]{0,0,0}\makebox(0,0)[b]{\smash{F2}}}%
    \put(0.95900088,0.70696492){\color[rgb]{0,0,0}\makebox(0,0)[b]{\smash{F3}}}%
    \put(0.53698853,0.57919332){\color[rgb]{0,0,0}\makebox(0,0)[b]{\smash{(a)}}}%
    \put(0,0){\includegraphics[width=\unitlength,page=2]{timing_concept1.pdf}}%
    \put(0.53698853,0.00815605){\color[rgb]{0,0,0}\makebox(0,0)[b]{\smash{(b)}}}%
    \put(0,0){\includegraphics[width=\unitlength,page=3]{timing_concept1.pdf}}%
    \put(0.17637879,0.7978802){\color[rgb]{0,0,0}\makebox(0,0)[b]{\smash{A}}}%
    \put(0.46187721,0.75941073){\color[rgb]{0,0,0}\makebox(0,0)[b]{\smash{B}}}%
    \put(0.60558293,0.76059049){\color[rgb]{0,0,0}\makebox(0,0)[b]{\smash{C}}}%
    \put(0.88840716,0.78079047){\color[rgb]{0,0,0}\makebox(0,0)[b]{\smash{D}}}%
    \put(0.01266991,0.39890492){\color[rgb]{0,0,0}\makebox(0,0)[b]{\smash{A}}}%
    \put(0.01205434,0.29547214){\color[rgb]{0,0,0}\makebox(0,0)[b]{\smash{B}}}%
    \put(0.1691971,0.42996592){\color[rgb]{0,0,0.50196078}\makebox(0,0)[lt]{\begin{minipage}{0.42746598\unitlength}\raggedright data\_a\_1\end{minipage}}}%
    \put(0.43837852,0.42996592){\color[rgb]{0,0.50196078,0.50196078}\makebox(0,0)[lt]{\begin{minipage}{0.42746598\unitlength}\raggedright data\_a\_ 2\end{minipage}}}%
    \put(0.01205434,0.19393277){\color[rgb]{0,0,0}\makebox(0,0)[b]{\smash{C}}}%
    \put(0.01205434,0.09112931){\color[rgb]{0,0,0}\makebox(0,0)[b]{\smash{D}}}%
    \put(0.44313126,0.22478225){\color[rgb]{0,0,0.50196078}\makebox(0,0)[lt]{\begin{minipage}{0.42746598\unitlength}\raggedright data\_c\_1\end{minipage}}}%
    \put(0.71293763,0.22478225){\color[rgb]{0,0.50196078,0.50196078}\makebox(0,0)[lt]{\begin{minipage}{0.42746598\unitlength}\raggedright data\_c\_2\end{minipage}}}%
    \put(0.55723215,0.12219045){\color[rgb]{0,0,0.50196078}\makebox(0,0)[lt]{\begin{minipage}{0.42746598\unitlength}\raggedright data\_d\_1\end{minipage}}}%
    \put(0.09934966,0.54829717){\color[rgb]{0,0,0}\makebox(0,0)[b]{\smash{0}}}%
    \put(0.37292801,0.55074002){\color[rgb]{0,0,0}\makebox(0,0)[b]{\smash{T}}}%
    \put(0.66018957,0.5502235){\color[rgb]{0,0,0}\makebox(0,0)[b]{\smash{2T}}}%
    \put(0.93533004,0.54953787){\color[rgb]{0,0,0}\makebox(0,0)[b]{\smash{3T}}}%
    \put(0,0){\includegraphics[width=\unitlength,page=4]{timing_concept1.pdf}}%
    \put(0.71410931,0.42996592){\color[rgb]{0.50196078,0.50196078,0.50196078}\makebox(0,0)[lt]{\begin{minipage}{0.42746598\unitlength}\raggedright data\_a\_3\end{minipage}}}%
    \put(0,0){\includegraphics[width=\unitlength,page=5]{timing_concept1.pdf}}%
    \put(0.07379035,0.36008972){\color[rgb]{0,0,0}\makebox(0,0)[lb]{\smash{$\small{t_{cq}}$}}}%
    \put(0,0){\includegraphics[width=\unitlength,page=6]{timing_concept1.pdf}}%
    \put(0.30437446,0.32737132){\color[rgb]{0,0,0.50196078}\makebox(0,0)[lt]{\begin{minipage}{0.42746598\unitlength}\raggedright data\_b\_1\end{minipage}}}%
    \put(0.57845982,0.32737132){\color[rgb]{0,0.50196078,0.50196078}\makebox(0,0)[lt]{\begin{minipage}{0.42746598\unitlength}\raggedright data\_b\_2\end{minipage}}}%
    \put(0.82365608,0.32737132){\color[rgb]{0.50196078,0.50196078,0.50196078}\makebox(0,0)[lt]{\begin{minipage}{0.42746598\unitlength}\raggedright data\_b\_3\end{minipage}}}%
    \put(0,0){\includegraphics[width=\unitlength,page=7]{timing_concept1.pdf}}%
    \put(0.83518859,0.12219045){\color[rgb]{0,0.50196078,0.50196078}\makebox(0,0)[lt]{\begin{minipage}{0.42746598\unitlength}\raggedright data\_d\_2\end{minipage}}}%
  \end{picture}%
\endgroup%

\caption{Conventional timing (a) A part of a sequential circuit. (b) Single-period clocking.}
\label{fig:timing_concept1}
}
\end{figure}

&  With the retiming technique, the area overhead to construct wave-pipelining paths 
can also be reduced, 
since paths not related to wave-pipelining construction are maintained as single-period 
to avoid unnecessary delay insertion.  

& A camouflaged netlist with wave-pipelining 
only contains normal logic gates, so that it is 
challenging for attackers to isolate and then identify the locations 
of wave-pipelining. %
An attack method based on path delay test to locate the camouflaged wave-pipelining paths 
would require a large number of test vectors.
& The introduced wave-pipelining false paths obstruct the test-based 
counterfeiting methods
further, because some paths that are originally testable are camouflaged as
false paths in the netlist.

& The proposed method is fully compatible with other security techniques 
introduced previously, so that they can be combined together to strengthen
netlist security %
at circuit level.

\end{easylist}

The rest of this paper is organized as follows. In
Section~\ref{sec:motivation}, we explain the motivation and the basic idea of the proposed
method. In Section~\ref{sec:wave_pipelining}, we provide a detailed description of
the wave-pipelining technique. 
In Section~\ref{sec:counter_measures}, we analyze potential attack
techniques to identify or circumvent wave-pipelining paths introduced into the
circuit. We also discuss 
the limitations of these techniques and propose countermeasures 
to thwart the attack attempts. We describe the implementation details to
construct wave-pipelining paths %
in Section~\ref{sec:impl}.
Experimental results are reported in Section~\ref{sec:results}.
Conclusions are drawn in Section~\ref{sec:conclusion}.

\section{Motivation and Basic Concept}\label{sec:motivation}

\begin{figure}[t]
{\figurefontsize
\centering
\begingroup%
  \makeatletter%
  \providecommand\color[2][]{%
    \errmessage{(Inkscape) Color is used for the text in Inkscape, but the package 'color.sty' is not loaded}%
    \renewcommand\color[2][]{}%
  }%
  \providecommand\transparent[1]{%
    \errmessage{(Inkscape) Transparency is used (non-zero) for the text in Inkscape, but the package 'transparent.sty' is not loaded}%
    \renewcommand\transparent[1]{}%
  }%
  \providecommand\rotatebox[2]{#2}%
  \ifx\svgwidth\undefined%
    \setlength{\unitlength}{219.87631095bp}%
    \ifx\svgscale\undefined%
      \relax%
    \else%
      \setlength{\unitlength}{\unitlength * \real{\svgscale}}%
    \fi%
  \else%
    \setlength{\unitlength}{\svgwidth}%
  \fi%
  \global\let\svgwidth\undefined%
  \global\let\svgscale\undefined%
  \makeatother%
  \begin{picture}(1,0.83267636)%
    \put(0,0){\includegraphics[width=\unitlength,page=1]{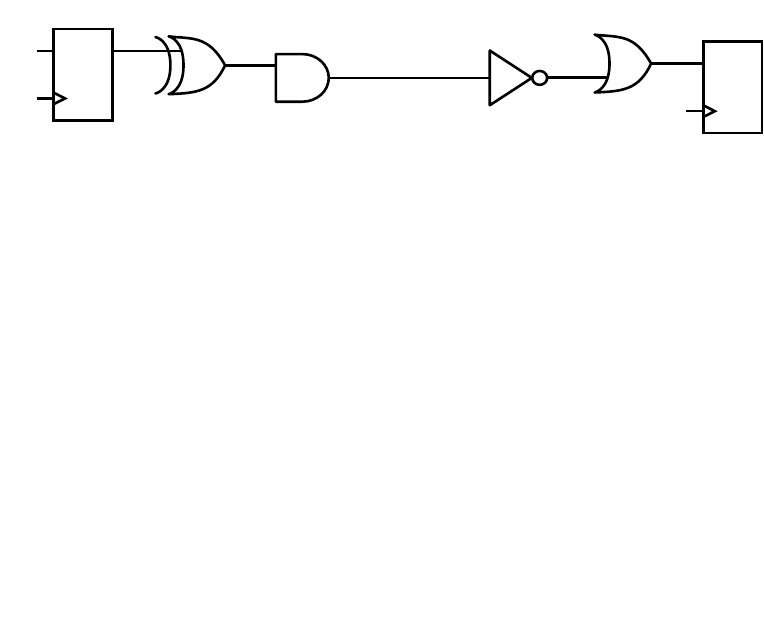}}%
    \put(0.11098672,0.72092556){\color[rgb]{0,0,0}\makebox(0,0)[b]{\smash{F1}}}%
    \put(0.95910541,0.70469351){\color[rgb]{0,0,0}\makebox(0,0)[b]{\smash{F3}}}%
    \put(0,0){\includegraphics[width=\unitlength,page=2]{timing_concept2.pdf}}%
    \put(0.17847867,0.79537698){\color[rgb]{0,0,0}\makebox(0,0)[b]{\smash{A}}}%
    \put(0.46324925,0.75700559){\color[rgb]{0,0,0}\makebox(0,0)[b]{\smash{B}}}%
    \put(0.60658857,0.75818235){\color[rgb]{0,0,0}\makebox(0,0)[b]{\smash{C}}}%
    \put(0.88869179,0.77833082){\color[rgb]{0,0,0}\makebox(0,0)[b]{\smash{D}}}%
    \put(0,0){\includegraphics[width=\unitlength,page=3]{timing_concept2.pdf}}%
    \put(0.21075107,0.68087623){\color[rgb]{0,0.50196078,0.50196078}\makebox(0,0)[b]{\smash{wave 2}}}%
    \put(0,0){\includegraphics[width=\unitlength,page=4]{timing_concept2.pdf}}%
    \put(0.54225861,0.67661173){\color[rgb]{0,0,0.50196078}\makebox(0,0)[b]{\smash{wave 1}}}%
    \put(0,0){\includegraphics[width=\unitlength,page=5]{timing_concept2.pdf}}%
    \put(0.53561942,0.00813526){\color[rgb]{0,0,0}\makebox(0,0)[b]{\smash{(b)}}}%
    \put(0.01263766,0.39917604){\color[rgb]{0,0,0}\makebox(0,0)[b]{\smash{A}}}%
    \put(0.16876569,0.43015782){\color[rgb]{0,0,0.50196078}\makebox(0,0)[lt]{\begin{minipage}{0.42637608\unitlength}\raggedright data\_a\_1\end{minipage}}}%
    \put(0.43726079,0.43015782){\color[rgb]{0,0.50196078,0.50196078}\makebox(0,0)[lt]{\begin{minipage}{0.42637608\unitlength}\raggedright data\_a\_ 2\end{minipage}}}%
    \put(0.30355685,0.32782756){\color[rgb]{0,0,0.50196078}\makebox(0,0)[lt]{\begin{minipage}{0.42637608\unitlength}\raggedright data\_b\_1\end{minipage}}}%
    \put(0.57694337,0.32782756){\color[rgb]{0,0.50196078,0.50196078}\makebox(0,0)[lt]{\begin{minipage}{0.42637608\unitlength}\raggedright data\_b\_2\end{minipage}}}%
    \put(0.01202361,0.29684578){\color[rgb]{0,0,0}\makebox(0,0)[b]{\smash{B}}}%
    \put(0.01202361,0.19472659){\color[rgb]{0,0,0}\makebox(0,0)[b]{\smash{C}}}%
    \put(0.01202361,0.09218506){\color[rgb]{0,0,0}\makebox(0,0)[b]{\smash{D}}}%
    \put(0.09909635,0.54818728){\color[rgb]{0,0,0}\makebox(0,0)[b]{\smash{0}}}%
    \put(0.37197721,0.550624){\color[rgb]{0,0,0}\makebox(0,0)[b]{\smash{T}}}%
    \put(0.65850624,0.5501088){\color[rgb]{0,0,0}\makebox(0,0)[b]{\smash{2T}}}%
    \put(0.93294529,0.54942482){\color[rgb]{0,0,0}\makebox(0,0)[b]{\smash{3T}}}%
    \put(0,0){\includegraphics[width=\unitlength,page=6]{timing_concept2.pdf}}%
    \put(0.71228857,0.43015782){\color[rgb]{0.50196078,0.50196078,0.50196078}\makebox(0,0)[lt]{\begin{minipage}{0.42637608\unitlength}\raggedright data\_a\_3\end{minipage}}}%
    \put(0.82151445,0.32782756){\color[rgb]{0.50196078,0.50196078,0.50196078}\makebox(0,0)[lt]{\begin{minipage}{0.42637608\unitlength}\raggedright data\_b\_3\end{minipage}}}%
    \put(0,0){\includegraphics[width=\unitlength,page=7]{timing_concept2.pdf}}%
    \put(0.30355679,0.22549737){\color[rgb]{0,0,0.50196078}\makebox(0,0)[lt]{\begin{minipage}{0.42637608\unitlength}\raggedright data\_c\_1\end{minipage}}}%
    \put(0.57694331,0.22549737){\color[rgb]{0,0.50196078,0.50196078}\makebox(0,0)[lt]{\begin{minipage}{0.42637608\unitlength}\raggedright data\_c\_2\end{minipage}}}%
    \put(0.8215144,0.22549737){\color[rgb]{0.50196078,0.50196078,0.50196078}\makebox(0,0)[lt]{\begin{minipage}{0.42637608\unitlength}\raggedright data\_c\_3\end{minipage}}}%
    \put(0,0){\includegraphics[width=\unitlength,page=8]{timing_concept2.pdf}}%
    \put(0.53561942,0.57528003){\color[rgb]{0,0,0}\makebox(0,0)[b]{\smash{(a)}}}%
    \put(0,0){\includegraphics[width=\unitlength,page=9]{timing_concept2.pdf}}%
    \put(0.52826172,0.12194808){\color[rgb]{0,0,0.50196078}\makebox(0,0)[lt]{\begin{minipage}{0.42637608\unitlength}\raggedright data\_d\_1\end{minipage}}}%
    \put(0.80115448,0.12194808){\color[rgb]{0,0.50196078,0.50196078}\makebox(0,0)[lt]{\begin{minipage}{0.42637608\unitlength}\raggedright data\_d\_2\end{minipage}}}%
    \put(0,0){\includegraphics[width=\unitlength,page=10]{timing_concept2.pdf}}%
  \end{picture}%
\endgroup%

\caption{Wave-pipelining. (a) The sequential circuit after F2 is removed. (b) Pipelining with two data waves.}
\label{fig:timing_concept2}
}
\end{figure}

Digital circuits rely on their structures to define their functions. A netlist
is usually sufficient to reproduce a correctly working circuit. 
To prevent the netlist from being recognized by reverse engineering, 
techniques from physical level to netlist level can be applied to 
camouflage the logic.
These methods, 
however, are still confined within the conventional single-period clocking timing model, so
that attackers 
only need to recognize the netlist correctly. 

In the conventional clocking timing model, all the paths in a combinational block
operate within one clock period. We call them \textbf{\textit{single-period clocking paths}}. 
Figure~\ref{fig:timing_concept1}(a) shows an example of a conventional 
sequential circuit with
three flip-flops F1, F2 and F3. 
The data switching activities at internal points A, B, C and D in this circuit 
are illustrated in Figure~\ref{fig:timing_concept1}(b). 
We assume that data are latched into flip-flops at the rising clock edge.  
At time 0, the input data of F1 is transferred to its output and 
becomes stable after $t_{cq}$, the clock-to-q delay of F1, shown as data\_a\_1.  It 
travels further 
through the logic gates and reaches B, 
shown as data\_b\_1 
in \figname~\ref{fig:timing_concept1}(b). 
Although the data at B is stable far before the next rising 
clock at time T, it is still blocked at F2 until the arrival of 
the next rising clock edge to be transferred to the output of F2. 
At this clock edge, 
the second data wave is injected onto the path from A to B by F1 and starts to propagate.      
In this way, combinational logic blocks are isolated by
flip-flops and data waves are pipelined to propagate through the logic blocks 
in the conventional digital design. %

A side effect of the sequential isolation with flip-flops above is that the
netlist carries all logic information. 
This simplification allows attackers to counterfeit chips relatively easily, 
because they only need to recognize the logic types of gates, flip-flops, 
and interconnect connections with reverse engineering.

\textit{To thwart the potential netlist 
attack attempt described above, we propose to invalidate the
conventional timing model in the circuit under protection.} 
For example, we can remove
the flip-flop in the middle of \figname~\ref{fig:timing_concept1}(a) to
construct the circuit structure shown in \figname~\ref{fig:timing_concept2}(a). 
The
switching activities of the internal signals in \figname~\ref{fig:timing_concept2}(a)
are illustrated in \figname~\ref{fig:timing_concept2}(b).
At two consecutive rising clock edges, F1
injects data\_a\_1 and data\_a\_2 onto the combinational path, respectively.
Therefore, two data waves at A are always separated by one clock period.
Since no flip-flop blocks the propagation of data\_b\_1,
it passes through C directly and reaches D after traveling through the inverter and OR gate.
Once the data at D becomes stable as data\_d\_1, it waits to be
latched by F3 while the second data wave is propagating.
To avoid that data\_d\_1 is flushed by the following
data waves,
the delays of all the combinational paths passing through B and C, including those between F1 and F3,  %
must 
be larger than one clock period. Otherwise,
the change of the data at D triggered by the second data wave data\_d\_2
happens before the next rising clock edge, so that 
the previous data\_d\_1 waiting at the input of F3 cannot be latched correctly. 
The result of flip-flop removal is that two data waves propagate along the path 
without a flip-flop separating them. 
This 
technique is called \textit{wave-pipelining (WP)} and has previously been investigated
for circuit optimization as in, e.g., \cite{Hsieh1995,Burleson1998,SeetharamanV09,Grace2018_DAC}.

With wave-pipelining paths, the function of the 
circuit depends on both its structure and the timing information of combinational paths. 
\textit{If 
attackers obtain a netlist as in \figname~\ref{fig:timing_concept2}(a), they
need to determine whether these paths are single-period clocking paths or wave-pipelining
paths.} If attackers assume the former and process the netlist using a
standard EDA flow, the circuit loses synchronization, because the 
data at the input of F3 is latched one clock period earlier than in the original design.  
If attackers want to determine whether it is the latter case, 
additional effort is required to
extract the timing information for each combinational path
in the circuit. 

Though attackers may have access to the standard cell library, e.g., 
through a third-party IP vendor,
it is still very hard to obtain accurate interconnect/RC parasitics 
by delayering authentic chips, due to unknown process parameters, 
challenges in 3D RC extraction, and switching-window-dependent 
crosstalk-induced delay variations, etc. 
In any case, the more accurate 
the original timing information should be recognized from delayered chips,
the harder and more expensive it becomes to reproduce a design. 
Therefore, this 
unconventional timing concept has a potential to open up a new 
dimension of netlist security and can be combined with the existing camouflaging techniques, 
 e.g., through dopant-level camouflage \cite{BeckerRPB14}, or dummy
contact insertion\cite{MalikBPB15,Rajendran2013,RajendranSK13}.   %

\section{Wave-Pipelining Constraints}\label{sec:wave_pipelining}

A wave-pipelining path such as the one in \figname~\ref{fig:timing_concept2}(a)
allows two data waves to propagate on the path simultaneously. 
Since the second data wave
must not catch the first one, special timing constraints should be imposed
for this path. %

When creating wave-pipelining paths, a flip-flop in \figname~\ref{fig:timing_concept1}(a) is removed 
to
construct the circuit in \figname~\ref{fig:timing_concept2}(a). In practice, this operation may lead to many
paths with wave pipelining, because any combinational path 
through B and C becomes a new wave-pipelining path. 
When the setup time $t_{su}$ and the hold time $t_{h}$ of the flip-flop are considered, 
all these paths $P$ 
should meet two constraints. First, the delay $d_p$ of a path should be $t_h$ larger than
the clock period $T$. Otherwise, the second wave arrives at the flip-flop 
too early and thus disturbs the latching process of the first wave. 
Second, the delay of the path should be $t_{su}$ smaller 
than $2T$ to 
guarantee that the data is latched by F3 correctly.
The timing constraints for all these
paths can be written as 
\begin{align} \label{eq:short_path}
d_p\ge T+t_h, \forall p\in P &\iff \min_{p\in P}\{d_p-t_h\}\ge T \\
\label{eq:long_path}
d_p\le 2T-t_{su}, \forall p\in P &\iff \max_{p\in P}\{d_p+t_{su}\}\le 2T.
\end{align}
If all the wave-pipelining paths
meet the two constraints (\ref{eq:short_path}) and (\ref{eq:long_path}), the
wave-pipelining version of the circuit after a flip-flop is removed 
is functionally equivalent to the
original circuit.

\section{\reviewhl{Potential Attacks and Countermeasures}}\label{sec:counter_measures}

The proposed camouflage technique with wave-pipelining secures netlists  
with timing information at sequential level. 
\reviewhl{This 
new technique may face potential attacks. We analyze some of these attacks in
this section, though their experimental attempt is not covered in this manuscript.
}
In the assumed attack model, the available information includes a netlist
recognized by reverse engineering and estimated delays of logic gates as well as
interconnects with an inaccuracy factor $\tau$. The objective of the attack is
to identify on which combinational paths in the netlist wave-pipelining is
applied.

\subsubsection{ First Attack Technique -- Delay Estimation}
In this method, the delays of all the gates and interconnects are measured while the netlist is
reverse engineered. Using the measured delays, path
delays can be estimated from the netlist. Since the delays of
wave-pipelining paths are between $T$ and $2T$ as defined in
(\ref{eq:short_path}) and (\ref{eq:long_path}),
these paths can therefore be identified.
The challenge of this attack technique is that it is difficult to extract
accurate gate and interconnect delays just from reverse engineering,  
due to the inaccuracy in delayering authentic chips described in Section~\ref{sec:motivation}.  
Assume that the real delay of a path is $d$, including setup time of a flip-flop, and
the delay recognition technique suffers an inaccuracy factor $\tau$ ($0<\tau<1$).
Consequently, this path delay can be any value in the range
[$(1-\tau)d$, $(1+\tau)d$]. If the upper bound of an estimated delay is smaller than
$T$, this path is definitely a single-period clocking path. If the lower bound
of an estimated delay is larger than $T$, the path is definitely a wave-pipelining path.
However, if 
no such clear decision can be made with the estimated delay, namely, 
\begin{equation}\label{eq:gray}
(1-\tau)d\le T\le (1+\tau)d 
\end{equation}
this path can only be considered as suspicious of wave-pipelining. 
In the following, we call the range
[$(1-\tau)d$, $(1+\tau)d$] the \textit{gray region} for a path with delay $d$.
In reality, a well-optimized design contains a huge number of 
critical paths with delays close to the clock period $T$, 
so that their gray regions often surround $T$.
When constructing
wave-pipelining paths in the proposed method, we also guarantee that their delays are in the gray
region to counter this attack technique.

\subsubsection{Second Attack Technique -- Testing Delays}

With the estimated delays, attackers can actually narrow down the number
of potential wave-pipelining paths, because paths
with estimated 
delays definitely smaller or larger than $T$ %
can be screened out.
The second attack technique is thus to
test the delays of the remaining suspicious paths using authentic chips from the market.
With the netlist recognized,
it is not difficult to determine test vectors to
trigger the remaining suspicious paths. Since the only information of interest
is whether a path delay is larger than $T$, only one delay test for
each path is required.

To prevent all suspicious paths from being tested as described above, 
we introduce a
countermeasure to create unsensitizable paths with wave-pipelining.
When we construct wave-pipelining paths by removing
flip-flops, we prefer the paths that, viewed directly with the conventional
single-period clocking model, 
are false paths, which cannot be sensitized
by any test vector.
\begin{mydef} 
False Path: A combinational path which cannot be activated in functional mode or tested due to controlling signals
from other paths\cite{Yuan2010,DuYG89}. %
\end{mydef}
\begin{mydef}
Wave-Pipelining False Path (WP False Path): A combinational path with wave pipelining that is a false path when viewed
with the conventional single-period clocking model.
\end{mydef}
\begin{mydef}
Wave-Pipelining True Path (WP True Path): A combinational path with wave pipelining that is a true path when viewed
with the conventional single-period clocking model.
\end{mydef}

Wave-pipelining false paths have two data waves propagating along them
when the circuit is operating, but they are false paths when the netlist is
examined assuming the 
conventional single-period clocking model. 
An example of a wave-pipelining false path 
is shown in \figname~\ref{fig:aa_false}, which is a snippet of
the s298 circuit from the ISCAS89 benchmark set.
When the flip-flop in the middle is removed, the dashed path becomes a wave-pipelining path. 
If attackers view it as a single-period clocking path in the extracted netlist, this 
dashed path is also a false path. 
In this case,
 a signal switching at the beginning of the dashed path never reaches the final flip-flop. If the signal $v_2$ has
a value `1', which is the controlling signal to an OR gate, it blocks the signal switching 
along the dashed path at the last OR gate;
If the signal $v_2$ has a value `0', it blocks the signal switching along 
the dashed path at the AND gate right away.
Consequently, the dashed path cannot
be triggered for delay test and attackers have no way to differentiate it from
all the other false paths in the original circuit, which may contribute up to 75\% of
all the combinational paths in real circuits\cite{Heragu1997}. 

\subsubsection{Third Attack Technique -- Logic Simulation}
Since the delays of false paths cannot be tested, %
brute-force logic simulation could be applied 
to differentiate the camouflaged false paths from real false paths. 
In this method, each false path that cannot be excluded by delay screening in
the first step can be assumed to be a real false path or a wave-pipelining false path,  
so that attackers have to verify which assumption is correct for this false path 
with simulations.   
Assuming the number of such paths is $n$.  
If 
each path in $n$ can be a real false path or a wave-pipelining false path, 
then $2^n$ simulations of the complete circuit have to be performed to check which
combination is correct. 
In theory, this method can eventually find the correct
combination of real false paths and wave-pipelining false paths.
However, it is still impractical because of the unaffordable simulation time
due to the large number of false paths in the original
design \cite{Heragu1997,Yuan2010}, e.g., 728262 for s13207 in the ISCAS89 benchmark set, 
 and the long runtime for a full
simulation of the complete circuit.

\subsubsection{Fourth Attack Technique -- Sizing False Paths}
In this method, all
false paths in the circuit are considered 
as wave-pipelining paths and logic gates are sized so
that delays of all these paths meet the constraints (\ref{eq:short_path}) and
(\ref{eq:long_path}). The concept behind this technique is that false paths are
not triggered anyway so that they do not affect the logic of the circuit
if their delays are larger than the clock period $T$.
This assumption, however, is incorrect because false paths
sized to have delays larger than $T$ may still affect the normal
circuit operation. Another challenge of this attack technique is
that it is very difficult to find a solution to size a huge number of false paths
without affecting the normal true paths whose delays should be
smaller than $T$.

\begin{figure}[t]
{\figurefontsize
\centering
\begingroup%
  \makeatletter%
  \providecommand\color[2][]{%
    \errmessage{(Inkscape) Color is used for the text in Inkscape, but the package 'color.sty' is not loaded}%
    \renewcommand\color[2][]{}%
  }%
  \providecommand\transparent[1]{%
    \errmessage{(Inkscape) Transparency is used (non-zero) for the text in Inkscape, but the package 'transparent.sty' is not loaded}%
    \renewcommand\transparent[1]{}%
  }%
  \providecommand\rotatebox[2]{#2}%
  \ifx\svgwidth\undefined%
    \setlength{\unitlength}{230.55974846bp}%
    \ifx\svgscale\undefined%
      \relax%
    \else%
      \setlength{\unitlength}{\unitlength * \real{\svgscale}}%
    \fi%
  \else%
    \setlength{\unitlength}{\svgwidth}%
  \fi%
  \global\let\svgwidth\undefined%
  \global\let\svgscale\undefined%
  \makeatother%
  \begin{picture}(1,0.30942051)%
    \put(0,0){\includegraphics[width=\unitlength,page=1]{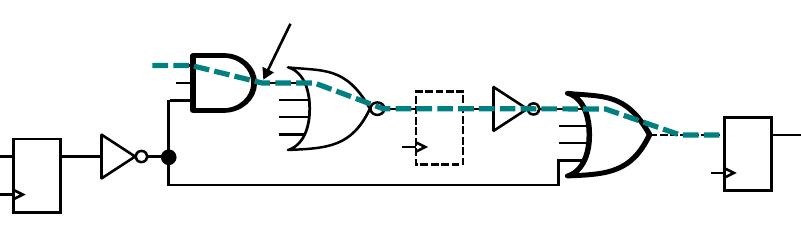}}%
    \put(0.34397003,0.28799005){\color[rgb]{0,0,0}\makebox(0,0)[b]{\smash{false path after wave-pipelining}}}%
    \put(0,0){\includegraphics[width=\unitlength,page=2]{aa_false.pdf}}%
    \put(0.52564218,0.00638229){\color[rgb]{0,0,0}\makebox(0,0)[b]{\smash{controlling signal}}}%
    \put(0,0){\includegraphics[width=\unitlength,page=3]{aa_false.pdf}}%
    \put(0.66168484,0.2426504){\color[rgb]{0,0,0}\makebox(0,0)[b]{\smash{removed flip-flop}}}%
    \put(0.10261166,0.12354518){\color[rgb]{0,0,0}\makebox(0,0)[b]{\smash{$v_1$}}}%
    \put(0.27086261,0.08741149){\color[rgb]{0,0,0}\makebox(0,0)[b]{\smash{$v_2$}}}%
    \put(0.04438082,0.00523364){\color[rgb]{0,0,0}\makebox(0,0)[b]{\smash{$F_1$}}}%
    \put(0.9381296,0.03369698){\color[rgb]{0,0,0}\makebox(0,0)[b]{\smash{$F_2$}}}%
  \end{picture}%
\endgroup%

\caption{Two true paths form a wave-pipelining false path.}
\label{fig:aa_false} 
}
\end{figure}

\subsubsection{Fifth Attack Technique -- Delay Calculation} 
In this method, all gate and interconnect delays in a circuit are calculated from path delays measured by testing. 
Since path delays
are linear combinations of gate and interconnect delays, the measured path delays can be used
to calculate gate and interconnect delays by linear algebra. 
This method needs to deal with several new challenges. 
First, in a commercial design, 
a large number of combinational paths will be tested. 
Second, all logic gates and interconnect segments 
should appear on testable paths in a way 
that the coefficient matrix of the system of linear equations has a rank equal
to the number of gates and interconnect segments, even in view of a large percentage of false
paths. Third, 
 inaccuracy in at-speed test of path delays exists 
due to environmental factors such as noise and temperature as well as 
the nature of binary-search of at-speed delay test.

\subsubsection{Sixth Attack Technique -- Pattern Recognition} 
To construct wave-pipelining paths such as the dashed path in \figname~\ref{fig:aa_false},
the flip-flop in the middle is removed,
leading to a relatively large number of logic gates along this path.
In addition, interconnects might be lengthened to enlarge 
the path delay to satisfy the timing constraints 
(\ref{eq:short_path}) and (\ref{eq:long_path}).
The special patterns of long interconnects and 
gate chains might expose the locations of wave-pipelining paths.
The sixth attack technique is to 
search such 
special patterns %
to identify the wave-pipelining paths. 
However, this attack method faces many challenges. 
First, the uncritical single-period paths have 
smaller delays compared with the critical paths. 
To guarantee the timing 
constraints of the critical paths,   
interconnects along them are usually short. 
On the contrary, the length of interconnects along uncritical paths might be large. 
Consequently, 
whether wave-pipelining exists cannot be determined simply by 
the length of interconnects. 
Second, the delays of the logic gates can be camouflaged with multi-threshold voltage (MVT)  
technique \cite{Multivt,Ruchir2004,Frank2005}, where high $V_t$ gates have larger delays than low $V_t$ gates. 
From the view of layout, there is no difference between both types of gates, 
 because 
the modulation of the threshold voltages is achieved by changing 
channel doping concentration during manufacturing \cite{Ithi2016}. 
To reduce the number of 
logic gates along wave-pipelining paths, 
the logic gates can be replaced 
with high $V_t$ counterparts to enlarge their delays. 
Consequently, 
attackers cannot determine the locations of wave-pipelining paths according 
to the number of logic gates. %
\subsubsection{Seventh Attack Technique -- SAT-based Attack}
In the SAT-based attack methods, e.g.,\cite{Meng2018,Abhi16}, 
it is assumed that attackers have full access to the scan chain, so that 
they can apply input test vectors and observe the outputs with the authentic chips. 
With 
various sets of input and output observations,   
the SAT-based attack can determine the 
locations of wave-pipelining paths that match all input and output 
observations. %
In fact, wave-pipelining true paths can be screened out with this method, 
since 
these paths can be triggered with the at-speed testing. 
However, the constructed wave-pipelining false paths cannot be triggered for 
delay test if they are considered to work within one clock period. 
If attackers try to activate such paths with two consecutive data waves, 
all the side-inputs of the wave-pipelining false paths %
should be set to non-controlling values in two consecutive clock cycles. 
For example, $v_2$ in \figname~\ref{fig:aa_false} is one of the side-inputs of 
the dashed wave-pipelining false path. It should be set to 1 
in the first clock cycle to 
allow a signal switching through the AND gate. 
In the second clock cycle, it should be set to 0 so that the signal switching can pass 
through the OR gate.   
These requirements might be met by tracing logic blocks between 
two flip-flop stages before F1.  
The traced logic blocks should be 
 set to appropriate values so that 
all the side-inputs ensure the activation of the wave-pipelining false paths. 
However, this method 
requires drastic changes in the existing testing platform.  
\hlreview{
In addition, the delays of original false paths in the circuits might be larger than the given clock period 
since they are ignored during timing analysis. 
By triggering the conflicting logic with two clock cycles, 
these paths can also be activated with two consecutive data waves, so that
the constructed wave-pipelining false paths can be concealed.
Furthermore, TimingCamouflage+ can be combined with other existing methods such
as scan chain encryption
\cite{Mathies2019,Byang2006}, 
to further increase the difficulty of
activating wave-pipelining paths.
}

\hlreview{
To differentiate wave-pipelining false paths from
original false paths with SAT-based attacks,
the whole circuit can be considered as a black box, where only
the data at
the primary inputs and the primary outputs of the design can be observed.
Since identifying wave-pipelining false paths 
requires to determine
where the flip-flops are removed,
attackers can first collect
connections between gates along all
suspicious false paths
and then determine where to re-insert flip-flops to recover the original
circuit without wave-pipelining.
To identify the correct combinations of inserting flip-flops,
SAT-based attacks search iteratively for discriminating input sequences at the
primary inputs.
Each
discriminating input sequence eliminates one or more combinations of inserting flip-flops.
The iteration 
continues until
only the correct combination of inserting flip-flops remains. 
In \cite{Mohamed2017,Kaveh2019}, 
similar attacks have been attempted onto sequential circuits, 
and it has been demonstrated that it is not possible to decamouflage relatively large 
sequential circuits even with smaller numbers of keys compared with
TimingCamouflage+.
Furthermore, TimingCamouflage+ actually introduces a new dimension in netlist
camouflaging. Therefore, it
can also be combined with other security methods, e.g., Anti-SAT 
logic locking\cite{antisat2019,Mitigat2016}, to counter SAT-based attacks
together. }

Recently, a Satisfiability Modulo Theory (SMT)-based attack method is proposed to 
enhance SAT-based attack with theory and graph solvers \cite{Azar2019SMTAN}. 
However, it is assumed that all combinational paths work within one clock period. 
This assumption does not hold in 
TimingCamouflage+, where the intentionally constructed wave-pipelining paths have delays
larger than one clock period. To extract correct timing constraints of such wave-pipelining paths, 
their locations have to be identified, which requires much effort as discussed
above.

\section{Wave-pipelining Construction}\label{sec:impl}

When constructing wave-pipelining paths in a circuit while maintaining its
original function, we need to guarantee that the constructed paths meet the
timing constraints (\ref{eq:short_path}) and (\ref{eq:long_path}). 
To counter
the attack techniques discussed in Section~\ref{sec:counter_measures}, 
the constructed paths %
should meet the constraint (\ref{eq:gray}), so that they cannot be verified easily.  
Furthermore, the wave-pipelining paths
should contain false paths when viewed as single-period clocking paths. 
The wave-pipelining construction problem can thus be formulated as follows.

\begin{easylist}
&\textit{Inputs}: Original optimized design; gate and interconnect delays; 
the given clock period $T$; the delay
recognition inaccuracy factor $\tau$ ($0<\tau<1$); the required number of 
wave-pipelining false and true paths $n_{wpf}$, $n_{wpt}$; distance threshold $dis_t$. 

&\textit{Outputs}: A revised design 
containing at least $n_{wpf}$ wave-pipelining false paths 
and $n_{wpt}$ wave-pipelining true paths, where $n_{wpf}$
and $n_{wpt}$ are user-defined parameters.
These wave-pipelining paths should meet the timing constraints 
(\ref{eq:short_path}) and (\ref{eq:long_path}) 
as well as 
the gray region
requirement (\ref{eq:gray}). %

&\textit{Objectives}: The original function of the circuit 
should be maintained; the original design should be kept unchanged as much as
possible; the increased resource usage should be as little as possible; 
the physical distances between the constructed wave-pipelining paths should be as far as possible to prevent the exposure of 
these paths resulting from clustering.  

\end{easylist}

\begin{figure}[t]
{\figurefontsize
\centering
\begingroup%
  \makeatletter%
  \providecommand\color[2][]{%
    \errmessage{(Inkscape) Color is used for the text in Inkscape, but the package 'color.sty' is not loaded}%
    \renewcommand\color[2][]{}%
  }%
  \providecommand\transparent[1]{%
    \errmessage{(Inkscape) Transparency is used (non-zero) for the text in Inkscape, but the package 'transparent.sty' is not loaded}%
    \renewcommand\transparent[1]{}%
  }%
  \providecommand\rotatebox[2]{#2}%
  \ifx\svgwidth\undefined%
    \setlength{\unitlength}{273.54776507bp}%
    \ifx\svgscale\undefined%
      \relax%
    \else%
      \setlength{\unitlength}{\unitlength * \real{\svgscale}}%
    \fi%
  \else%
    \setlength{\unitlength}{\svgwidth}%
  \fi%
  \global\let\svgwidth\undefined%
  \global\let\svgscale\undefined%
  \makeatother%
  \begin{picture}(1,0.7212346)%
    \put(0.01697286,0.68265075){\color[rgb]{0,0,0}\makebox(0,0)[lb]{\smash{Input: netlist, delay information, $T$, $\tau$, $n_{wpf}$, $n_{wpt}$, $dis_t$; }}}%
    \put(0,0){\includegraphics[width=\unitlength,page=1]{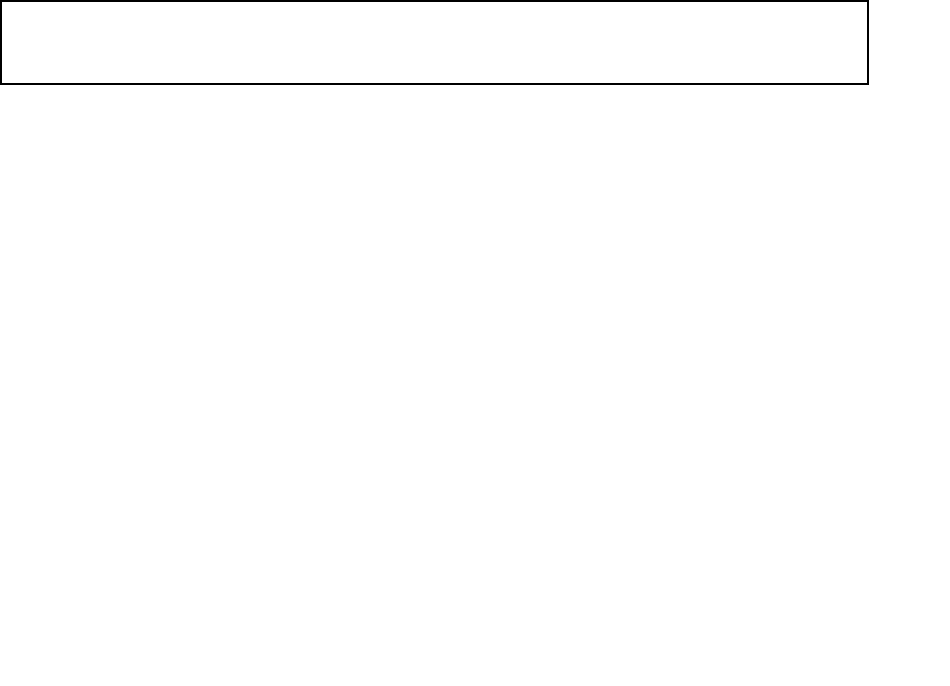}}%
    \put(0.01697286,0.48620207){\color[rgb]{0,0,0}\makebox(0,0)[lb]{\smash{For i=1 to $|$F$_n$$|$do}}}%
    \put(0.14283025,0.41271751){\color[rgb]{0,0,0}\makebox(0,0)[lb]{\smash{$n_f$=check\_WP\_false\_paths($\mathit{ff_i}$, $T$, $\tau$);}}}%
    \put(0.14283025,0.37595546){\color[rgb]{0,0,0}\makebox(0,0)[lb]{\smash{If $n_f>0$ then}}}%
    \put(0.20586204,0.33919342){\color[rgb]{0,0,0}\makebox(0,0)[lb]{\smash{If construct\_WP\_paths($\mathit{ff_i}$, $T$, $\tau$)=false, go to L5;}}}%
    \put(0.14283025,0.1553832){\color[rgb]{0,0,0}\makebox(0,0)[lb]{\smash{If $n_{wpf}\le 0$ then}}}%
    \put(0.14225856,0.19214524){\color[rgb]{0,0,0}\makebox(0,0)[lb]{\smash{$n_{wpf}=n_{wpf}-n_f$;}}}%
    \put(0.14333633,0.11772362){\color[rgb]{0,0,0}\makebox(0,0)[lb]{\smash{break;}}}%
    \put(0,0){\includegraphics[width=\unitlength,page=2]{work_flow.pdf}}%
    \put(0.01697286,0.64670516){\color[rgb]{0,0,0}\makebox(0,0)[lb]{\smash{F$_w=\varnothing$;}}}%
    \put(0.07990157,0.44947955){\color[rgb]{0,0,0}\makebox(0,0)[lb]{\smash{If a flip-flop $\mathit{ff_i}\notin$F$_w$  then}}}%
    \put(0.20575895,0.30243137){\color[rgb]{0,0,0}\makebox(0,0)[lb]{\smash{F$_w$$\gets$ $\mathit{ff_i}$, fanin($\mathit{ff_i}$) and fanout($\mathit{ff_i}$);}}}%
    \put(0,0){\includegraphics[width=\unitlength,page=3]{work_flow.pdf}}%
    \put(0.02043914,0.01673107){\color[rgb]{0,0,0}\makebox(0,0)[lb]{\smash{Construct wave-pipelining true paths similar to L5--L15}}}%
    \put(0.84790768,0.68133557){\color[rgb]{0,0,0}\makebox(0,0)[lb]{\smash{L1}}}%
    \put(0.84790768,0.41075987){\color[rgb]{0,0,0}\makebox(0,0)[lb]{\smash{L7}}}%
    \put(0.84790768,0.15343457){\color[rgb]{0,0,0}\makebox(0,0)[lb]{\smash{L14}}}%
    \put(0.84790768,0.4844138){\color[rgb]{0,0,0}\makebox(0,0)[lb]{\smash{L5}}}%
    \put(0.84790768,0.11667252){\color[rgb]{0,0,0}\makebox(0,0)[lb]{\smash{L15}}}%
    \put(0.84790768,0.1900668){\color[rgb]{0,0,0}\makebox(0,0)[lb]{\smash{L13}}}%
    \put(0.20575895,0.26566933){\color[rgb]{0,0,0.05882353}\makebox(0,0)[lb]{\smash{If$(DIS(\mathit{ff_i},\mathit{ff_j})<dis_t)$}}}%
    \put(0.01697286,0.55964244){\color[rgb]{0,0,0}\makebox(0,0)[lb]{\smash{Sort all flip-flops F in a decreasing order;}}}%
    \put(0.01697286,0.52300364){\color[rgb]{0,0,0}\makebox(0,0)[lb]{\smash{Filter F $\to$ F$_n$ using the number of source/sink flip-flops; }}}%
    \put(0.84790768,0.55781694){\color[rgb]{0,0,0}\makebox(0,0)[lb]{\smash{L3}}}%
    \put(0.84790768,0.52092526){\color[rgb]{0,0,0}\makebox(0,0)[lb]{\smash{L4}}}%
    \put(0.84790768,0.22682885){\color[rgb]{0,0,0}\makebox(0,0)[lb]{\smash{L12}}}%
    \put(0.84790768,0.2635909){\color[rgb]{0,0,0}\makebox(0,0)[lb]{\smash{}}}%
    \put(0.2765219,0.22890729){\color[rgb]{0,0,0}\makebox(0,0)[lb]{\smash{F$_w$$\gets$ $\mathit{ff_j}$;}}}%
    \put(0.84790768,0.33724483){\color[rgb]{0,0,0}\makebox(0,0)[lb]{\smash{L9}}}%
    \put(0.84790768,0.30048278){\color[rgb]{0,0,0}\makebox(0,0)[lb]{\smash{L10}}}%
    \put(0.14742707,0.64549646){\color[rgb]{0,0,0}\makebox(0,0)[lb]{\smash{// the flip-flops that cannot be used for construction}}}%
    \put(0.52277321,0.26635631){\color[rgb]{0,0,0}\makebox(0,0)[lb]{\smash{//$\mathit{ff_j}$ are other flip-flops}}}%
    \put(0.74328601,0.264745){\color[rgb]{0,0,0}\makebox(0,0)[lb]{\smash{}}}%
  \end{picture}%
\endgroup%

\caption{Major steps of wave-pipelining construction.}
\label{fig:work_flow}
}
\end{figure}

When constructing wave-pipelining paths, we incorporate PVT (Process, Voltage and Temperature) variations by
allowing path delays to deviate from their original values. %
Specifically, delays of longest paths are enlarged to (1+$\delta$) times of the original values. 
On the contrary, delays of shortest paths are reduced to (1-$\delta$) times of the original values. 
The value of $\delta$ should be determined by designers according to 
the corresponding manufacturing technology. 
With this setting, the constructed wave-pipelining 
paths should still work correctly under PVT 
variations.

\subsection{Work flow of wave-pipelining construction}\label{sec:workflow}

The major steps to construct wave-pipelining paths are shown in
\figname~\ref{fig:work_flow}. 
Wave-pipelining paths can potentially be  
constructed 
at a flip-flop 
connected to paths with large delays, 
so that  
timing constraints (\ref{eq:short_path}) and (\ref{eq:long_path}) 
of such paths 
can be satisfied easily. 
Therefore, we sort all flip-flops in a circuit in a decreasing order 
according to 
the sum of the maximum delays 
of their incoming and outgoing paths as described above.  %
In addition, 
wave-pipelining construction might be challenging 
at those flip-flops with a large number of incoming 
and outgoing paths, because 
a huge number of wave-pipelining paths can appear when the flip-flop is removed. 
To guarantee all these wave-pipelining paths to 
meet the timing constraints (\ref{eq:short_path}) and (\ref{eq:long_path}) 
is difficult, because these constraints require that all the path delays should be within the range of T and 2T simultaneously. 
Therefore, such flip-flops with large numbers of incoming and outgoing paths should be filtered out.  
Since traversing all incoming and outgoing paths of a flip-flop to 
acquire their numbers     
is time-consuming, we use the numbers of source and sink flip-flops of a flip-flop
to indicate the difficulty in constructing wave-pipelining paths.  
Therefore, 
flip-flops with the number of source or sink flip-flops larger than a given a threshold are 
filtered out to accelerate the construction.  %

After sorting and filtering flip-flops, 
for each remaining flip-flop $\mathit{ff_i}$, we check whether there are wave-pipelining
false paths that can be formed from single-period true paths on the left and on the
right of 
$\mathit{ff_i}$ (L7).  
The number of such paths 
is stored in $n_f$. 
If wave-pipelining false paths can be formed at $\mathit{ff_i}$, the function
construct\_WP\_paths($\mathit{ff_i}$, $T$, $\tau$) is used to construct such
paths eventually with 
the combinational logic leaving from and arriving at it. 
The identified logic gates are also expanded
to include all the gates reachable from them, 
because sizing the logic gates on the wave-pipelining
paths also affects delays of paths through the expanded gates. All these gates
are denoted together as a set $G$.
The 
details of this construction will be explained later. 

As shown in \figname~\ref{fig:timing_concept2}(a), for a wave-pipelining path, 
 the flip-flop at the beginning
of the path and the flip-flop at the end of 
the path should not be removed from the circuit during constructing of 
further wave-pipelining paths. 
Otherwise, paths with more than 2 waves may appear, requiring more complex timing constraints.  
These fanin and fanout flip-flops 
are inserted into the
set F$_w$ (L10) 
and all the flip-flops tracked by F$_w$ cannot be considered as
candidates to construct wave-pipelining paths. %
In addition, the physical distance between the wave-pipelining paths 
should be large  %
to prevent the exposure resulting from the clustering of such paths.  
We use the distances between flip-flops to represent the distances between wave-pipelining paths. 
Therefore, a flip-flop $\mathit{ff_j}$ whose distance to $\mathit{ff_i}$ that is currently used to construct wave-pipelining 
is smaller than $dis_t$
cannot be a candidate 
for wave-pipelining, so that it is inserted into the set F$_w$. %
\begin{figure}[t]
{\figurefontsize
\centering
\begingroup%
  \makeatletter%
  \providecommand\color[2][]{%
    \errmessage{(Inkscape) Color is used for the text in Inkscape, but the package 'color.sty' is not loaded}%
    \renewcommand\color[2][]{}%
  }%
  \providecommand\transparent[1]{%
    \errmessage{(Inkscape) Transparency is used (non-zero) for the text in Inkscape, but the package 'transparent.sty' is not loaded}%
    \renewcommand\transparent[1]{}%
  }%
  \providecommand\rotatebox[2]{#2}%
  \ifx\svgwidth\undefined%
    \setlength{\unitlength}{207.42218674bp}%
    \ifx\svgscale\undefined%
      \relax%
    \else%
      \setlength{\unitlength}{\unitlength * \real{\svgscale}}%
    \fi%
  \else%
    \setlength{\unitlength}{\svgwidth}%
  \fi%
  \global\let\svgwidth\undefined%
  \global\let\svgscale\undefined%
  \makeatother%
  \begin{picture}(1,0.30800928)%
    \put(0.03703806,0.00036158){\color[rgb]{0,0,0}\makebox(0,0)[b]{\smash{fanin}}}%
    \put(0.95375375,0.00036158){\color[rgb]{0,0,0}\makebox(0,0)[b]{\smash{fanout}}}%
    \put(0,0){\includegraphics[width=\unitlength,page=1]{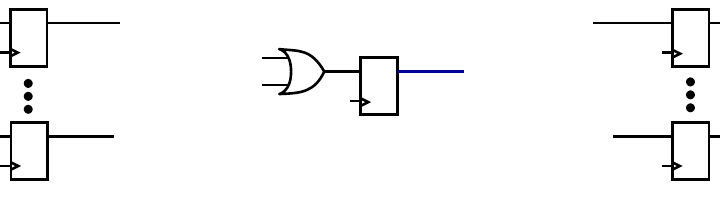}}%
    \put(0.25844417,0.0510191){\color[rgb]{0,0,0}\makebox(0,0)[b]{\smash{500 path limit}}}%
    \put(0.74054674,0.05244741){\color[rgb]{0,0,0}\makebox(0,0)[b]{\smash{500 path limit}}}%
    \put(0,0){\includegraphics[width=\unitlength,page=2]{500path.pdf}}%
    \put(0.22108404,0.26858882){\color[rgb]{0,0,0}\makebox(0,0)[b]{\smash{F}}}%
    \put(0.17270406,0.23581098){\color[rgb]{0,0,0}\makebox(0,0)[b]{\smash{T}}}%
    \put(0.17782321,0.14453468){\color[rgb]{0,0,0}\makebox(0,0)[b]{\smash{F}}}%
    \put(0.21752562,0.10660569){\color[rgb]{0,0,0}\makebox(0,0)[b]{\smash{T}}}%
    \put(0,0){\includegraphics[width=\unitlength,page=3]{500path.pdf}}%
    \put(0.75309904,0.19754912){\color[rgb]{0,0,0}\makebox(0,0)[b]{\smash{F}}}%
    \put(0.70924505,0.22960477){\color[rgb]{0,0,0}\makebox(0,0)[b]{\smash{T}}}%
    \put(0.72058186,0.1477047){\color[rgb]{0,0,0}\makebox(0,0)[b]{\smash{T}}}%
    \put(0,0){\includegraphics[width=\unitlength,page=4]{500path.pdf}}%
    \put(0.52624319,0.1784273){\color[rgb]{0,0,0}\makebox(0,0)[b]{\smash{$\mathit{ff_i}$}}}%
  \end{picture}%
\endgroup%

\caption{The number of paths on each side of $\mathit{ff_i}$ is limited to 500.}
\vspace{-1.2em}
\label{fig:500path}
}
\end{figure}

In the last step of the proposed  method, we construct wave-pipelining 
paths that are still true viewed with the single-period clocking model.
These paths are used to guarantee that attackers must test all 
single-period clocking and wave-pipelining true paths
whose delays meet the gray region requirement.
Without these paths, attackers can assume all testable paths are 
single-period and avoid the expensive test procedure. 
Different from the construction of wave-pipelining false paths,
the construction of these true paths only relies on path delays.
The path
construction in this step is similar to  L5--L15 in
\figname~\ref{fig:work_flow}. The only differences in this construction are that at L7 we 
check wave-pipelining true paths and in L13 and L14 we use $n_{wpt}$ as the number
of such paths to be constructed.

\subsection{False path checking}\label{sec:false_checking}

In the work flow above, we need to check very often 
whether a path is false or not. 
In the proposed method, we only consider the statically unsensitizable paths as
false paths \cite{Devadas2006,Coudert2010},
such as the false path shown in 
\figname~\ref{fig:aa_false}. In this example, the path cannot be sensitized
because the controlling signal $v_2$ blocks either the AND gate or the last OR gate, 
no matter what its value is. 
Besides statically unsensitizable paths, there are also dynamically
unsensitizable paths \cite{Yuan2010},
which, however, might still be sensitized by test vectors set through the scan chain \cite{Kim2003}. 
Therefore, 
statically unsensitizable paths are more conservative in thwarting attacks and 
thus used in our method.

To verify whether a path is statically unsensitizable, we assign Boolean
variables to the inputs and output of each gate
and formulate false path checking as a 
SAT
problem \cite{Coudert2010}. 
The logic relations between these variables are established according to 
functions of the corresponding logic gates.
If a path can be sensitized, all the side inputs of the path
must be set to the non-controlling values. For example, the path in
\figname~\ref{fig:aa_false} requires that the condition 
$(v_2 \land \lnot v_2)$ is true, which is, however, always false.

In implementing the function
check\_WP\_false\_paths($\mathit{\mathit{ff_i}}$, $T$, $\tau$) in
\figname~\ref{fig:work_flow}, we randomly select 500 paths that drive the current
flip-flop $\mathit{ff_i}$ and exclude the false paths from them, 
because the wave-pipelining paths to be constructed should be formed 
by two single-period clocking true paths.
Similarly, we select 500 paths that are driven by $\mathit{ff_i}$ and exclude the false
paths. 
By limiting the number of paths we avoid to enumerate all paths
while searching wave-pipelining false paths, which are in fact abundant in the
circuits as demonstrated by experimental results in Section~\ref{sec:results}.
The concept of this path selection is illustrated in
\figname~\ref{fig:500path}.
  
\subsection{Retiming to facilitate wave-pipelining construction}\label{sec:retiming}

\begin{figure}[t]
{\figurefontsize
\centering
\begingroup%
  \makeatletter%
  \providecommand\color[2][]{%
    \errmessage{(Inkscape) Color is used for the text in Inkscape, but the package 'color.sty' is not loaded}%
    \renewcommand\color[2][]{}%
  }%
  \providecommand\transparent[1]{%
    \errmessage{(Inkscape) Transparency is used (non-zero) for the text in Inkscape, but the package 'transparent.sty' is not loaded}%
    \renewcommand\transparent[1]{}%
  }%
  \providecommand\rotatebox[2]{#2}%
  \ifx\svgwidth\undefined%
    \setlength{\unitlength}{207.67798176bp}%
    \ifx\svgscale\undefined%
      \relax%
    \else%
      \setlength{\unitlength}{\unitlength * \real{\svgscale}}%
    \fi%
  \else%
    \setlength{\unitlength}{\svgwidth}%
  \fi%
  \global\let\svgwidth\undefined%
  \global\let\svgscale\undefined%
  \makeatother%
  \begin{picture}(1,0.66506813)%
    \put(0.03699244,0.31931417){\color[rgb]{0,0,0}\makebox(0,0)[b]{\smash{fanin}}}%
    \put(0.95257886,0.31931417){\color[rgb]{0,0,0}\makebox(0,0)[b]{\smash{fanout}}}%
    \put(0,0){\includegraphics[width=\unitlength,page=1]{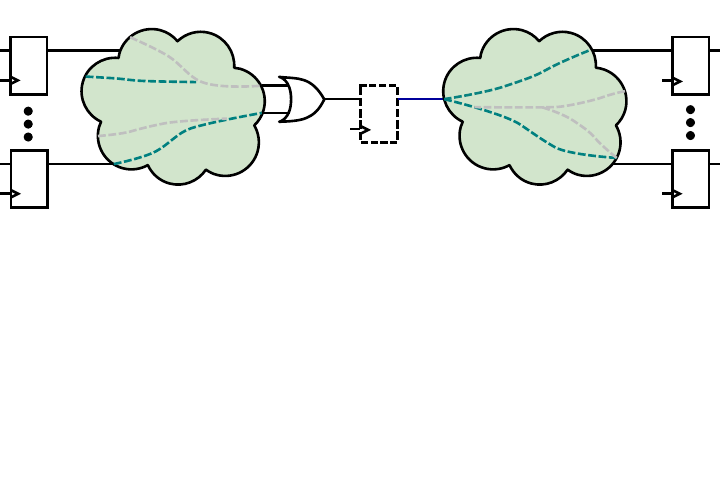}}%
    \put(0.52559497,0.49715942){\color[rgb]{0,0,0}\makebox(0,0)[b]{\smash{$\mathit{ff_i}$}}}%
    \put(0,0){\includegraphics[width=\unitlength,page=2]{relevant_paths.pdf}}%
    \put(0.15321056,0.53951412){\color[rgb]{0,0,0}\makebox(0,0)[lt]{\begin{minipage}{0.36936253\unitlength}\raggedright relevant \end{minipage}}}%
    \put(0.72059153,0.55494489){\color[rgb]{0,0,0}\makebox(0,0)[lt]{\begin{minipage}{0.36936253\unitlength}\raggedright relevant \end{minipage}}}%
    \put(0,0){\includegraphics[width=\unitlength,page=3]{relevant_paths.pdf}}%
    \put(0.42415445,0.66511583){\color[rgb]{0,0,0}\makebox(0,0)[lt]{\begin{minipage}{0.52364722\unitlength}\raggedright wave-pipelining false path\end{minipage}}}%
    \put(0,0){\includegraphics[width=\unitlength,page=4]{relevant_paths.pdf}}%
    \put(0.17608695,0.66867717){\color[rgb]{0,0,0}\makebox(0,0)[lt]{\begin{minipage}{0.36936253\unitlength}\raggedright irrelevant \end{minipage}}}%
    \put(0.14618718,0.39879246){\color[rgb]{0,0,0}\makebox(0,0)[lt]{\begin{minipage}{0.36936253\unitlength}\raggedright irrelevant \end{minipage}}}%
    \put(0,0){\includegraphics[width=\unitlength,page=5]{relevant_paths.pdf}}%
    \put(0.66675938,0.39879246){\color[rgb]{0,0,0}\makebox(0,0)[lt]{\begin{minipage}{0.36936253\unitlength}\raggedright irrelevant \end{minipage}}}%
    \put(0,0){\includegraphics[width=\unitlength,page=6]{relevant_paths.pdf}}%
    \put(0.50063194,0.0057529){\color[rgb]{0,0,0}\makebox(0,0)[b]{\smash{(b)}}}%
    \put(0,0){\includegraphics[width=\unitlength,page=7]{relevant_paths.pdf}}%
    \put(0.15321056,0.20796312){\color[rgb]{0,0,0}\makebox(0,0)[lt]{\begin{minipage}{0.36936253\unitlength}\raggedright relevant \end{minipage}}}%
    \put(0.4146739,0.20346666){\color[rgb]{0,0,0}\makebox(0,0)[b]{\smash{$\mathit{ff_j}$}}}%
    \put(0.4146739,0.09806914){\color[rgb]{0,0,0}\makebox(0,0)[b]{\smash{$\mathit{ff_k}$}}}%
    \put(0.72059153,0.21309059){\color[rgb]{0,0,0}\makebox(0,0)[lt]{\begin{minipage}{0.36936253\unitlength}\raggedright relevant \end{minipage}}}%
    \put(0.50063194,0.30885065){\color[rgb]{0,0,0}\makebox(0,0)[b]{\smash{(a)}}}%
    \put(0,0){\includegraphics[width=\unitlength,page=8]{relevant_paths.pdf}}%
    \put(0.17608695,0.31788262){\color[rgb]{0,0,0}\makebox(0,0)[lt]{\begin{minipage}{0.36936253\unitlength}\raggedright irrelevant \end{minipage}}}%
    \put(0.14618718,0.04654868){\color[rgb]{0,0,0}\makebox(0,0)[lt]{\begin{minipage}{0.36936253\unitlength}\raggedright irrelevant \end{minipage}}}%
    \put(0,0){\includegraphics[width=\unitlength,page=9]{relevant_paths.pdf}}%
    \put(0.66675938,0.05090902){\color[rgb]{0,0,0}\makebox(0,0)[lt]{\begin{minipage}{0.36936253\unitlength}\raggedright irrelevant \end{minipage}}}%
  \end{picture}%
\endgroup%

\caption{Removal and retiming of a flip-flop. (a) The generation of a wave-pipelining false path with the relevant paths and unnecessary wave-pipelining paths with the irrelevant paths after the flip-flop $\mathit{ff_i}$ is removed. (b) $\mathit{ff_i}$ is retimed to the left of the OR gate to block irrelevant paths.}
\label{fig:relevant_paths}
}
\end{figure}

With the false path checking method described above, we can check whether wave-pipelining
false paths can be formed from the selected single-period true paths 
on the left and right of flip-flop  
$\mathit{ff_i}$.  %
If such wave-pipelining false paths can be formed, 
we mark the original single-period clocking true paths forming them as \textbf{\textit{relevant paths}}. 
Other singe-period clocking paths are called \textbf{\textit{irrelevant paths}}. 
An example of the relevant and irrelevant paths is shown in \figname~\ref{fig:relevant_paths}(a).  
To construct wave-pipelining with relevant paths, 
the flip-flop $\mathit{ff_i}$ in the middle can be removed. 
Unfortunately, 
the removal of $\mathit{ff_i}$ makes all the paths connected with it wave-pipelining. 
Since many short paths may exist on the left and on the right of $\mathit{ff_i}$, 
connecting them by removing $\mathit{ff_i}$ 
directly generates many paths whose delays are too small to
meet the lower bound of the wave-pipelining constraint (\ref{eq:short_path}).
In addition, the removal of the flip-flop leads to
a clustering of wave-pipelining paths, which are vulnerable to be identified by attackers. 
\figname~\ref{fig:relevant_paths}(a) illustrates a construction example, 
where a wave-pipelining false path is formed from the relevant paths after $\mathit{ff_i}$ 
is removed. However, the irrelevant paths on the left and the right of $\mathit{ff_i}$ 
lead to unnecessary wave-pipelining generation. 
To deal with the challenges, 
we try to maintain the irrelevant paths %
on the left and right of $\mathit{ff_i}$ 
to be single-period as much as possible.

To maintain the irrelevant paths as single-period clocking paths, 
we apply the 
retiming technique \cite{Char91} to 
block these paths with flip-flops. %
Retiming transforms the structure of a circuit by 
moving the locations of flip-flops 
while preserving the function of this circuit. 
This concept is illustrated in \figname~\ref{fig:relevant_paths}(b), where 
the flip-flop $\mathit{ff_i}$ %
is moved to 
the left of the OR gate. 
To maintain the original function of the circuit, 
a flip-flop is inserted at each input of the OR gate.

\begin{figure}[t]
{\figurefontsize
\centering
\begingroup%
  \makeatletter%
  \providecommand\color[2][]{%
    \errmessage{(Inkscape) Color is used for the text in Inkscape, but the package 'color.sty' is not loaded}%
    \renewcommand\color[2][]{}%
  }%
  \providecommand\transparent[1]{%
    \errmessage{(Inkscape) Transparency is used (non-zero) for the text in Inkscape, but the package 'transparent.sty' is not loaded}%
    \renewcommand\transparent[1]{}%
  }%
  \providecommand\rotatebox[2]{#2}%
  \ifx\svgwidth\undefined%
    \setlength{\unitlength}{229.53039349bp}%
    \ifx\svgscale\undefined%
      \relax%
    \else%
      \setlength{\unitlength}{\unitlength * \real{\svgscale}}%
    \fi%
  \else%
    \setlength{\unitlength}{\svgwidth}%
  \fi%
  \global\let\svgwidth\undefined%
  \global\let\svgscale\undefined%
  \makeatother%
  \begin{picture}(1,0.58642617)%
    \put(0.5007787,0.00520519){\color[rgb]{0,0,0}\makebox(0,0)[b]{\smash{(b)}}}%
    \put(0,0){\includegraphics[width=\unitlength,page=1]{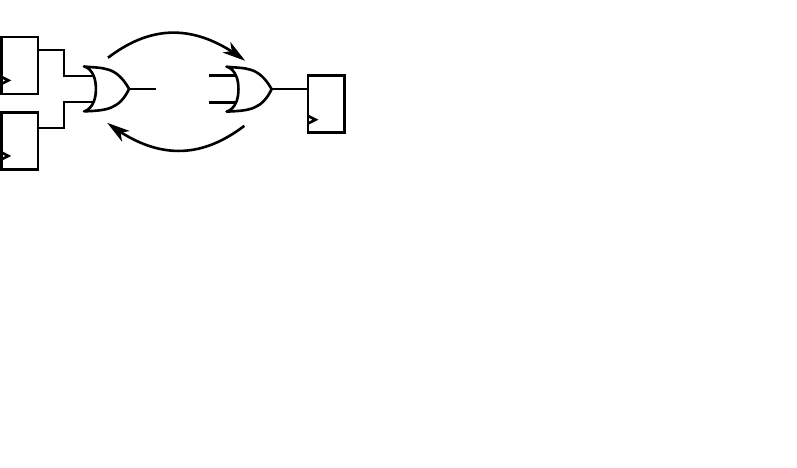}}%
    \put(0.12920218,0.58969331){\color[rgb]{0,0,0}\makebox(0,0)[lt]{\begin{minipage}{0.40103687\unitlength}\raggedright Retimed by -1\end{minipage}}}%
    \put(0.13392479,0.38800928){\color[rgb]{0,0,0}\makebox(0,0)[lt]{\begin{minipage}{0.40103687\unitlength}\raggedright Retimed by 1\end{minipage}}}%
    \put(0.21591342,0.32075672){\color[rgb]{0,0,0}\makebox(0,0)[b]{\smash{(a)}}}%
    \put(0.12636934,0.48303858){\color[rgb]{0,0,0}\makebox(0,0)[lt]{\begin{minipage}{0.13070164\unitlength}\raggedright $g$\end{minipage}}}%
    \put(0.30625917,0.48303858){\color[rgb]{0,0,0}\makebox(0,0)[lt]{\begin{minipage}{0.13070164\unitlength}\raggedright $g$\end{minipage}}}%
    \put(0.89628842,0.13917211){\color[rgb]{0,0,0}\makebox(0,0)[lb]{\smash{}}}%
    \put(0,0){\includegraphics[width=\unitlength,page=2]{retiming2_new.pdf}}%
    \put(0.47499624,0.34097826){\color[rgb]{0,0,0}\makebox(0,0)[b]{\smash{1}}}%
    \put(0,0){\includegraphics[width=\unitlength,page=3]{retiming2_new.pdf}}%
    \put(0.59376676,0.43098469){\color[rgb]{0,0,0}\makebox(0,0)[lt]{\begin{minipage}{0.26735791\unitlength}\centering $w(e_{g_i,g_j})$=0\end{minipage}}}%
    \put(0,0){\includegraphics[width=\unitlength,page=4]{retiming2_new.pdf}}%
    \put(0.65959417,0.48294973){\color[rgb]{0,0,0}\makebox(0,0)[lt]{\begin{minipage}{0.13070164\unitlength}\raggedright $g_i$\end{minipage}}}%
    \put(0,0){\includegraphics[width=\unitlength,page=5]{retiming2_new.pdf}}%
    \put(0.81406044,0.5004676){\color[rgb]{0,0,0}\makebox(0,0)[lt]{\begin{minipage}{0.11973542\unitlength}\raggedright $g_j$\end{minipage}}}%
    \put(0,0){\includegraphics[width=\unitlength,page=6]{retiming2_new.pdf}}%
    \put(0.4661272,0.08774379){\color[rgb]{0,0,0}\makebox(0,0)[b]{\smash{2}}}%
    \put(0,0){\includegraphics[width=\unitlength,page=7]{retiming2_new.pdf}}%
    \put(0.59477291,0.15870267){\color[rgb]{0,0,0}\makebox(0,0)[lt]{\begin{minipage}{0.13070164\unitlength}\raggedright $g_i$\end{minipage}}}%
    \put(0,0){\includegraphics[width=\unitlength,page=8]{retiming2_new.pdf}}%
    \put(0.87947479,0.17668735){\color[rgb]{0,0,0}\makebox(0,0)[lt]{\begin{minipage}{0.11973542\unitlength}\raggedright $g_j$\end{minipage}}}%
    \put(0,0){\includegraphics[width=\unitlength,page=9]{retiming2_new.pdf}}%
    \put(0.54103968,0.22943831){\color[rgb]{0,0,0}\makebox(0,0)[lt]{\begin{minipage}{0.40103687\unitlength}\raggedright $r(g_i)$=-1\end{minipage}}}%
    \put(0.86248043,0.22943831){\color[rgb]{0,0,0}\makebox(0,0)[lt]{\begin{minipage}{0.40103687\unitlength}\raggedright $r(g_j)$=1\end{minipage}}}%
    \put(0.67839537,0.07983145){\color[rgb]{0,0,0}\makebox(0,0)[lt]{\begin{minipage}{0.42478034\unitlength}\raggedright $w_r(e_{g_i,g_j})$=2\end{minipage}}}%
    \put(0.08868308,0.22943834){\color[rgb]{0,0,0}\makebox(0,0)[lt]{\begin{minipage}{0.40103687\unitlength}\raggedright $r(g_i)$=0\end{minipage}}}%
    \put(0.12788472,0.0808175){\color[rgb]{0,0,0}\makebox(0,0)[lt]{\begin{minipage}{0.42478034\unitlength}\raggedright $w_r(e_{g_i,g_j})$=1\end{minipage}}}%
    \put(0,0){\includegraphics[width=\unitlength,page=10]{retiming2_new.pdf}}%
    \put(0.12369371,0.15796472){\color[rgb]{0,0,0}\makebox(0,0)[lt]{\begin{minipage}{0.13070164\unitlength}\raggedright $g_i$\end{minipage}}}%
    \put(0,0){\includegraphics[width=\unitlength,page=11]{retiming2_new.pdf}}%
    \put(0.32997463,0.17594934){\color[rgb]{0,0,0}\makebox(0,0)[lt]{\begin{minipage}{0.11973542\unitlength}\raggedright $g_j$\end{minipage}}}%
    \put(0,0){\includegraphics[width=\unitlength,page=12]{retiming2_new.pdf}}%
    \put(0.31663752,0.22943834){\color[rgb]{0,0,0}\makebox(0,0)[lt]{\begin{minipage}{0.40103687\unitlength}\raggedright $r(g_j)$=1\end{minipage}}}%
    \put(0,0){\includegraphics[width=\unitlength,page=13]{retiming2_new.pdf}}%
  \end{picture}%
\endgroup%

\caption{Operations of retiming. (a) Basic operations of retiming. (b) Two retiming cases.}
\label{fig:retiming2}
}
\end{figure}

The concept of retiming can be explained using \figname~\ref{fig:retiming2}. %
To apply this technique, 
we use $g \in G$ to represent a combinational gate and a net $e \in E$ 
to represent the net connecting the output of a combinational gate and an 
input of another combinational gate. 
\hlreview{
The delays from the input pins of a gate to its output pin are different, due
to the internal structure of the gate.  Therefore, these
delays should be set individually according to the corresponding lookup table. 
Since the input pins of a gate appear along different combinational paths,
we use $d^p_g$ to represent the pin-to-pin delay for gate $g$ along the path $p$.} 
Interconnect delays can be modeled as extra nodes between nets, similar to combinational gates. 
Each net $e_{g_i,g_j}$ between gates $g_i$ and $g_j$ 
has a weight $w(e_{g_i,g_j})$ to represent the number of flip-flops along 
the connection. 
Assume that between two gates $g_i$ and $g_j$, %
there is a path $p$.   
The propagation delay of this path is equal to  
the sum of the delays of the gates on the path, expressed as follows 
\hlreview{
\begin{align}\label{eq:path_delay}
d(p) &= \sum_{k=1}^{n} d^p_{g_k}
\end{align}}
where $n$ is the number of gates on the path.  
Furthermore, the weight of a path $p$, representing the total number of flip-flops on the path, 
is defined as the sum of the weights of the 
nets along the path, %
expressed as 
\begin{align}\label{eq:path_weight}
w(p) &= \sum_{k=1}^{n-1} w(e_k).
\end{align}

The goal of retiming is to find an assignment of an integer $r(g)$ 
for each gate $g$ to transform a circuit to another functionally equivalent circuit. 
 $r(g)$ defines how many flip-flops are moved from the 
output of a gate to its inputs. 
In \figname~\ref{fig:retiming2}(a), 
the combinational gate is retimed by -1 if the flip-flops at its inputs are moved
to its output. In this case, the integer $r(g)$ of the gate is
equal to -1 . On the contrary,
the integer $r(g)$ of
the gate is 1, if the flip-flop at its output is moved to its inputs.

After retiming, the number of flip-flops on a net between gates $g_i$ and $g_j$ 
is written as $w_r(e_{g_i,g_j})=w(e_{g_i,g_j})+r(g_j)-r(g_i)$. 
Two retiming cases are shown in \figname~\ref{fig:retiming2}(b).
In the first case \ding{192}, the flip-flop at the output of $g_j$ is moved to its inputs,
so that the number of flip-flops between $g_i$ and $g_j$ can be derived as $w_r(e_{g_i,g_j})=w(e_{g_i,g_j})+r(g_j)-r(g_i)= 
0+1-0 = 1$.
In the second case \ding{193}, the flip-flops at the inputs of $g_i$ are moved to its output further,
so that
the number of flip-flops between $g_i$ and $g_j$ is increased to $w_r(e_{g_i,g_j})=w(e_{g_i,g_j})+r(g_j)-r(g_i)= 
0+1-(-1) = 2$. 
For a legal retiming,
the retimed weight $w_r(e_{g_i,g_j})$ must be non-negative.
To
meet a given clock period $T$ for a retimed circuit,
any path $p$ with a delay larger than the given clock period $T$ 
should have a retimed weight $w_r(p)$ larger than 0 to guarantee there is a flip-flop on it, 
so that this path does not affect the clock period.

To construct wave-pipelining paths, retiming 
can be applied to move the locations of flip-flops to block irrelevant paths, 
so that these paths can still be maintained as single-period to 
avoid unnecessary wave-pipelining generation.   
With the retimed flip-flops, the irrelevant paths that do not contribute 
to the construction of wave-pipelining paths still work within one clock period. 
The relevant paths are used to construct wave-pipelining paths by removing the retimed flip-flops. 
The details of this construction are explained in 
the following sections.

\subsection{Wave-pipelining construction with removal of flip-flops combined with retiming}\label{sec:wave_cons_removal}

\begin{figure}[t]
{\figurefontsize
\centering
\begingroup%
  \makeatletter%
  \providecommand\color[2][]{%
    \errmessage{(Inkscape) Color is used for the text in Inkscape, but the package 'color.sty' is not loaded}%
    \renewcommand\color[2][]{}%
  }%
  \providecommand\transparent[1]{%
    \errmessage{(Inkscape) Transparency is used (non-zero) for the text in Inkscape, but the package 'transparent.sty' is not loaded}%
    \renewcommand\transparent[1]{}%
  }%
  \providecommand\rotatebox[2]{#2}%
  \ifx\svgwidth\undefined%
    \setlength{\unitlength}{207.42219374bp}%
    \ifx\svgscale\undefined%
      \relax%
    \else%
      \setlength{\unitlength}{\unitlength * \real{\svgscale}}%
    \fi%
  \else%
    \setlength{\unitlength}{\svgwidth}%
  \fi%
  \global\let\svgwidth\undefined%
  \global\let\svgscale\undefined%
  \makeatother%
  \begin{picture}(1,0.63609273)%
    \put(0,0){\includegraphics[width=\unitlength,page=1]{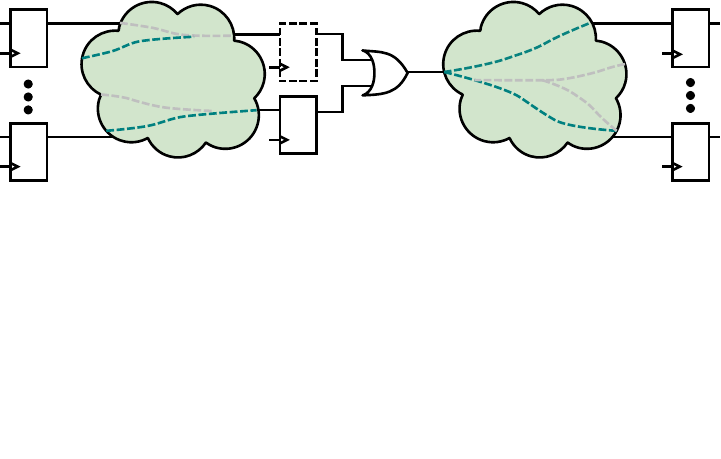}}%
    \put(0.50001608,0.36018499){\color[rgb]{0,0,0}\makebox(0,0)[b]{\smash{(a)}}}%
    \put(0.18467277,0.57710564){\color[rgb]{0,0,0}\makebox(0,0)[lt]{\begin{minipage}{0.36981802\unitlength}\raggedright relevant \end{minipage}}}%
    \put(0.35147269,0.64548283){\color[rgb]{0,0,0}\makebox(0,0)[lt]{\begin{minipage}{0.11098411\unitlength}\centering remove\end{minipage}}}%
    \put(0.41472337,0.55814257){\color[rgb]{0,0,0}\makebox(0,0)[b]{\smash{$\mathit{ff_j}$}}}%
    \put(0.41395203,0.45261487){\color[rgb]{0,0,0}\makebox(0,0)[b]{\smash{$\mathit{ff_k}$}}}%
    \put(0,0){\includegraphics[width=\unitlength,page=2]{retiming1.pdf}}%
    \put(0.50001608,0.04036677){\color[rgb]{0,0,0}\makebox(0,0)[b]{\smash{(b)}}}%
    \put(0.18467277,0.25728755){\color[rgb]{0,0,0}\makebox(0,0)[lt]{\begin{minipage}{0.36981802\unitlength}\raggedright relevant \end{minipage}}}%
    \put(0.41395203,0.13279686){\color[rgb]{0,0,0}\makebox(0,0)[b]{\smash{$\mathit{ff_k}$}}}%
    \put(0,0){\includegraphics[width=\unitlength,page=3]{retiming1.pdf}}%
    \put(0.36146218,0.34640739){\color[rgb]{0,0,0}\makebox(0,0)[lt]{\begin{minipage}{0.36981802\unitlength}\raggedright removal  \end{minipage}}}%
    \put(0.03703811,0.00036158){\color[rgb]{0,0,0}\makebox(0,0)[b]{\smash{fanin}}}%
    \put(0.95375366,0.00036158){\color[rgb]{0,0,0}\makebox(0,0)[b]{\smash{fanout}}}%
    \put(0.73167149,0.56821368){\color[rgb]{0,0,0}\makebox(0,0)[lt]{\begin{minipage}{0.36981802\unitlength}\raggedright relevant \end{minipage}}}%
    \put(0.38845366,0.31251145){\color[rgb]{0,0,0}\makebox(0,0)[lt]{\begin{minipage}{0.36981802\unitlength}\raggedright point\end{minipage}}}%
    \put(1.69317786,-5.01497691){\color[rgb]{0,0,0}\makebox(0,0)[lt]{\begin{minipage}{4.57813256\unitlength}\raggedright \end{minipage}}}%
    \put(0.73167149,0.2486547){\color[rgb]{0,0,0}\makebox(0,0)[lt]{\begin{minipage}{0.36981802\unitlength}\raggedright relevant \end{minipage}}}%
    \put(0,0){\includegraphics[width=\unitlength,page=4]{retiming1.pdf}}%
    \put(0.56521378,0.07135749){\color[rgb]{0,0,0}\makebox(0,0)[lt]{\begin{minipage}{0.38353456\unitlength}\raggedright wave-pipelining paths\end{minipage}}}%
    \put(0,0){\includegraphics[width=\unitlength,page=5]{retiming1.pdf}}%
    \put(0.83575412,0.31542479){\color[rgb]{0,0,0}\makebox(0,0)[lt]{\begin{minipage}{0.09620648\unitlength}\raggedright A\end{minipage}}}%
    \put(0.88139673,0.24653307){\color[rgb]{0,0,0}\makebox(0,0)[lt]{\begin{minipage}{0.09620648\unitlength}\raggedright B\end{minipage}}}%
    \put(0.87368645,0.15557766){\color[rgb]{0,0,0}\makebox(0,0)[lt]{\begin{minipage}{0.09620648\unitlength}\raggedright C\end{minipage}}}%
  \end{picture}%
\endgroup%

\caption{Removal of flip-flops together with retiming. (a) $\mathit{ff_j}$ is removed for wave-pipelining construction. (b) Wave-pipelining paths can be constructed from the relevant paths and a small number of irrelevant paths after the retimed flip-flop $\mathit{ff_j}$ is removed.}
\label{fig:retiming1}
}
\end{figure}

Retiming can facilitate the wave-pipelining construction   
by maintaining irrelevant paths as single-period as described above.   
We then  
construct wave-pipelining paths by removing 
flip-flops combined with retiming. 
A simple construction example is illustrated in \figname~\ref{fig:retiming1}(a), 
where the retimed flip-flop $\mathit{ff_j}$ is removed for wave-pipelining and the 
retimed flip-flop $\mathit{ff_k}$ is kept to block irrelevant paths. 
However, the removal of the retimed flip-flop $\mathit{ff_j}$ 
generates many wave-pipelining paths. 
To enlarge the delays of these paths to meet the lower bound defined in (\ref{eq:short_path}), 
a large area overhead might be incurred. 
Therefore, it is not straightforward   
to determine the optimal location to apply retiming to construct wave-pipelining. 
To deal with this challenge, 
we formulate the wave-pipelining construction %
as an ILP problem. 
The goal of the formulation is to achieve the construction of wave-pipelining paths 
whose delays meet the timing constraints (\ref{eq:short_path})--(\ref{eq:long_path}) 
as well as the gray region requirement (\ref{eq:gray}) with a minimum area overhead. 

\begin{figure}[t]
{\figurefontsize
\centering
\begingroup%
  \makeatletter%
  \providecommand\color[2][]{%
    \errmessage{(Inkscape) Color is used for the text in Inkscape, but the package 'color.sty' is not loaded}%
    \renewcommand\color[2][]{}%
  }%
  \providecommand\transparent[1]{%
    \errmessage{(Inkscape) Transparency is used (non-zero) for the text in Inkscape, but the package 'transparent.sty' is not loaded}%
    \renewcommand\transparent[1]{}%
  }%
  \providecommand\rotatebox[2]{#2}%
  \ifx\svgwidth\undefined%
    \setlength{\unitlength}{244.45925815bp}%
    \ifx\svgscale\undefined%
      \relax%
    \else%
      \setlength{\unitlength}{\unitlength * \real{\svgscale}}%
    \fi%
  \else%
    \setlength{\unitlength}{\svgwidth}%
  \fi%
  \global\let\svgwidth\undefined%
  \global\let\svgscale\undefined%
  \makeatother%
  \begin{picture}(1,0.183546)%
    \put(0,0){\includegraphics[width=\unitlength,page=1]{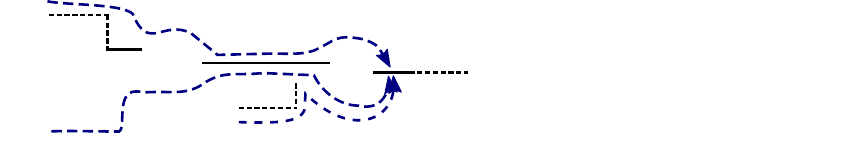}}%
    \put(0.28731363,0.13069376){\color[rgb]{0,0,0}\makebox(0,0)[lb]{\smash{$d_1$}}}%
    \put(0.28731363,0.01328441){\color[rgb]{0,0,0}\makebox(0,0)[lb]{\smash{$d_3$}}}%
    \put(0.28731363,0.07138768){\color[rgb]{0,0,0}\makebox(0,0)[lb]{\smash{$d_2$}}}%
    \put(0.63141771,0.13822936){\color[rgb]{0,0,0}\makebox(0,0)[lt]{\begin{minipage}{0.62858383\unitlength}\raggedright $\overline{s}_{g_j}= \max\{d_1,d_2,d_3\}$\end{minipage}}}%
    \put(0.63141771,0.08508255){\color[rgb]{0,0,0}\makebox(0,0)[lt]{\begin{minipage}{0.62858383\unitlength}\raggedright $\underline{s}_{g_j}= \min\{d_1,d_2,d_3\}$\end{minipage}}}%
    \put(0,0){\includegraphics[width=\unitlength,page=2]{arrival_times_gate.pdf}}%
    \put(0.20390614,0.1163933){\color[rgb]{0,0,0}\makebox(0,0)[lt]{\begin{minipage}{0.12271983\unitlength}\raggedright $g_i$\end{minipage}}}%
    \put(0,0){\includegraphics[width=\unitlength,page=3]{arrival_times_gate.pdf}}%
    \put(0.40830141,0.11021228){\color[rgb]{0,0,0}\makebox(0,0)[lt]{\begin{minipage}{0.1124233\unitlength}\raggedright $g_j$\end{minipage}}}%
    \put(0,0){\includegraphics[width=\unitlength,page=4]{arrival_times_gate.pdf}}%
    \put(0.47774126,0.11518758){\color[rgb]{0,0,0}\makebox(0,0)[lb]{\smash{$\overline{s}_{g_j}$}}}%
    \put(0.47774126,0.07222414){\color[rgb]{0,0,0}\makebox(0,0)[lb]{\smash{$\underline{s}_{g_j}$}}}%
  \end{picture}%
\endgroup%

\caption{The latest and the earliest arrival times of $g_j$.}
\label{fig:arrival_times}
}
\end{figure}

The removal of a retimed flip-flop leads to
many wave-pipelining paths from left and the right of the removal point, as shown
in \figname~\ref{fig:retiming1}(b). 
To guarantee the timing constraints (\ref{eq:short_path})--(\ref{eq:gray}),  
all wave-pipelining paths may be traversed to check their delays.  
However, this straightforward traversal %
might be time-consuming due to the number of paths. %
To accelerate this process, we assign two variables, 
$\overline{s}_g$ and $\underline{s}_g$ to represent 
the latest and the earliest arrival times 
at the output of a combinational gate $g$, as shown in \figname~\ref{fig:arrival_times}. 
The latest arrival time at gate $g_j$ is the maximum delay 
with which the data from a flip-flop travels through the longest 
path and arrives at the output of this gate, denoted as $\overline{s}_{g_j}=\max\{d_1,d_2,d_3\}$.  
On the contrary, 
the earliest arrival time at gate $g_j$ is the minimum delay
with which the data from a flip-flop travels through the shortest
path and arrives at the output of this gate, denoted as $\underline{s}_{g_j}=\min\{d_1,d_2,d_3\}$.  
To tolerate PVT variations, 
For the wave-pipelining paths after a flip-flop is removed, shown in \figname~\ref{fig:retiming1}(b), 
if the latest and the earliest arrival times at points A, B, C   
satisfy the timing constraints (\ref{eq:short_path})--(\ref{eq:long_path})
and the gray region requirement (\ref{eq:gray}), 
all the other wave-pipelining paths resulting from the removal of $\mathit{ff_j}$ are also guaranteed, 
because the arrival times of other paths are bounded between them.

Since we do not know at which location the retiming should be performed to construct %
wave-pipelining paths, %
a variable $y_{e_{g_i,g_j}}$ is assigned for each net $e_{g_i,g_j}$ %
to indicate whether the retimed flip-flop on 
this net can be removed, with $y_{e_{g_i,g_j}}=1$ to indicate that the retimed flip-flop is removed and vice versa.  
If there is no retimed flip-flop on $e_{g_i,g_j}$, $y_{e_{g_i,g_j}}$ should be set to 0. 
Consequently, the relation between $y_{e_{g_i,g_j}}$ and the number of retimed flip-flops $w_r(e_{g_i,g_j})$ 
on this net can be established as follows
\begin{align} 
      y_{e_{g_i,g_j}}\le w_r(e_{g_i,g_j}). \label{eq:remove_ff}
\end{align}
With this setting, 
three representative cases for a net between gates $g_i$ and $g_j$ %
should be examined, 
as shown in \figname~\ref{fig:shift_arrival}. 
Detailed timing constraints of each case are explained in Appendix.

\textbf{Case 1:} The net from gate $g_i$ to gate $g_j$ has the retimed weight $w_r(e_{g_i,g_j})=w(e_{g_i,g_j})+r(g_j)-r(g_i)= 1$, 
and thus a retimed flip-flop $\mathit{ff_{k_1}} $exists along this net. The retimed flip-flop is not removed,  %
denoted as $y_{e_{g_i,g_j}}=0$.

\textbf{Case 2:} %
The net from gate $g_i$ to $g_j$ has the retimed weight $w_r(e_{g_i,g_j})=w(e_{g_i,g_j})+r(g_j)-r(g_i)=1$, 
but the flip-flop is removed, denoted as $y_{e_{g_i,g_j}}=1$.  %

\begin{figure}[t]
{\figurefontsize
\centering
\begingroup%
  \makeatletter%
  \providecommand\color[2][]{%
    \errmessage{(Inkscape) Color is used for the text in Inkscape, but the package 'color.sty' is not loaded}%
    \renewcommand\color[2][]{}%
  }%
  \providecommand\transparent[1]{%
    \errmessage{(Inkscape) Transparency is used (non-zero) for the text in Inkscape, but the package 'transparent.sty' is not loaded}%
    \renewcommand\transparent[1]{}%
  }%
  \providecommand\rotatebox[2]{#2}%
  \ifx\svgwidth\undefined%
    \setlength{\unitlength}{235.07494003bp}%
    \ifx\svgscale\undefined%
      \relax%
    \else%
      \setlength{\unitlength}{\unitlength * \real{\svgscale}}%
    \fi%
  \else%
    \setlength{\unitlength}{\svgwidth}%
  \fi%
  \global\let\svgwidth\undefined%
  \global\let\svgscale\undefined%
  \makeatother%
  \begin{picture}(1,0.41163775)%
    \put(0,0){\includegraphics[width=\unitlength,page=1]{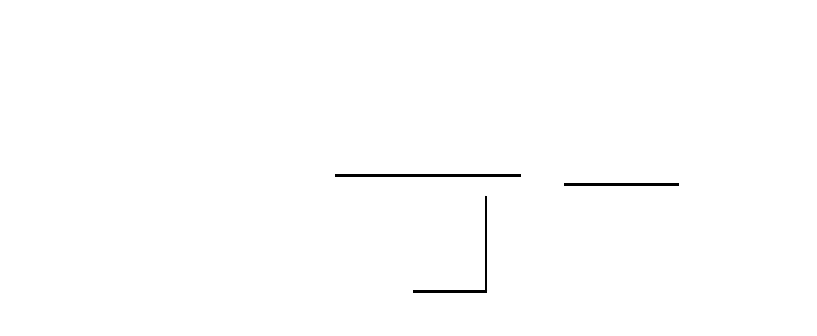}}%
    \put(0.30633582,0.31763062){\color[rgb]{0,0,0}\makebox(0,0)[lt]{\begin{minipage}{0.16111883\unitlength}\raggedright removed?\end{minipage}}}%
    \put(0,0){\includegraphics[width=\unitlength,page=2]{shift_arrival_gate.pdf}}%
    \put(0.47340294,0.3537513){\color[rgb]{0,0,0}\makebox(0,0)[lt]{\begin{minipage}{0.25444013\unitlength}\raggedright delay\end{minipage}}}%
    \put(0.472406,0.31763062){\color[rgb]{0,0,0}\makebox(0,0)[lt]{\begin{minipage}{0.16111883\unitlength}\raggedright inserted?\end{minipage}}}%
    \put(0.6325278,0.3537513){\color[rgb]{0,0,0}\makebox(0,0)[lt]{\begin{minipage}{0.25444013\unitlength}\raggedright gate\end{minipage}}}%
    \put(0.6325527,0.31763062){\color[rgb]{0,0,0}\makebox(0,0)[lt]{\begin{minipage}{0.16111883\unitlength}\raggedright sizing?\end{minipage}}}%
    \put(0.15245964,0.13794354){\color[rgb]{0,0,0}\makebox(0,0)[lt]{\begin{minipage}{0.1412923\unitlength}\centering $r(g_i)$\end{minipage}}}%
    \put(0.30802908,0.13794353){\color[rgb]{0,0,0}\makebox(0,0)[lt]{\begin{minipage}{0.31904719\unitlength}\raggedright $w_r(e_{v_i,v_j})$\end{minipage}}}%
    \put(0.30633576,0.3537513){\color[rgb]{0,0,0}\makebox(0,0)[lt]{\begin{minipage}{0.25444013\unitlength}\raggedright retimed ff\end{minipage}}}%
    \put(0.30653523,0.28138502){\color[rgb]{0,0,0}\makebox(0,0)[lt]{\begin{minipage}{0.35095191\unitlength}\raggedright $y_{e_{g_i,g_j}}$\end{minipage}}}%
    \put(0.63181746,0.28081529){\color[rgb]{0,0,0}\makebox(0,0)[lt]{\begin{minipage}{0.35095191\unitlength}\raggedright $d_{g_j}$\end{minipage}}}%
    \put(0.47269255,0.28195473){\color[rgb]{0,0,0}\makebox(0,0)[lt]{\begin{minipage}{0.35095191\unitlength}\raggedright $\xi_{g_j}$\end{minipage}}}%
    \put(0,0){\includegraphics[width=\unitlength,page=3]{shift_arrival_gate.pdf}}%
    \put(0.65742575,0.19789659){\color[rgb]{0,0,0}\makebox(0,0)[lt]{\begin{minipage}{0.1169113\unitlength}\raggedright $g_j$\end{minipage}}}%
    \put(0,0){\includegraphics[width=\unitlength,page=4]{shift_arrival_gate.pdf}}%
    \put(0.21453284,0.20965084){\color[rgb]{0,0,0}\makebox(0,0)[lt]{\begin{minipage}{0.12761888\unitlength}\raggedright $g_i$\end{minipage}}}%
    \put(2.89465787,-2.21664243){\color[rgb]{0,0,0}\makebox(0,0)[lt]{\begin{minipage}{0.03887688\unitlength}\raggedright \end{minipage}}}%
    \put(0.36044106,0.19570595){\color[rgb]{0,0,0}\makebox(0,0)[lt]{\begin{minipage}{0.40285793\unitlength}\raggedright $\mathit{ff_{k_1}}$\end{minipage}}}%
    \put(0.45499946,0.05146127){\color[rgb]{0,0,0}\makebox(0,0)[lt]{\begin{minipage}{0.40285793\unitlength}\raggedright $\mathit{ff_{k_2}}$\end{minipage}}}%
    \put(0.6079924,0.13794354){\color[rgb]{0,0,0}\makebox(0,0)[lt]{\begin{minipage}{0.1412923\unitlength}\centering $r(g_j)$\end{minipage}}}%
    \put(0.71489122,0.22390602){\color[rgb]{0,0,0}\makebox(0,0)[lt]{\begin{minipage}{0.19175264\unitlength}\raggedright $\overline{s}_{g_j}$\end{minipage}}}%
    \put(0.38938209,-0.02103162){\color[rgb]{0,0,0}\makebox(0,0)[lt]{\begin{minipage}{0.58011538\unitlength}\raggedright \end{minipage}}}%
    \put(0.71488903,0.17549831){\color[rgb]{0,0,0}\makebox(0,0)[lt]{\begin{minipage}{0.21764256\unitlength}\raggedright $\underline{s}_{g_j}$\end{minipage}}}%
    \put(0.25837982,0.23577294){\color[rgb]{0,0,0}\makebox(0,0)[lt]{\begin{minipage}{0.2828063\unitlength}\raggedright $\overline{s}_{g_i}$\end{minipage}}}%
    \put(0.25837766,0.18494758){\color[rgb]{0,0,0}\makebox(0,0)[lt]{\begin{minipage}{0.61153829\unitlength}\raggedright $\underline{s}_{g_i}$\end{minipage}}}%
    \put(0,0){\includegraphics[width=\unitlength,page=5]{shift_arrival_gate.pdf}}%
    \put(-0.00058439,0.08612864){\color[rgb]{0,0,0}\makebox(0,0)[lt]{\begin{minipage}{0.09990876\unitlength}\raggedright fanin\end{minipage}}}%
    \put(-0.00058814,0.05315902){\color[rgb]{0,0,0}\makebox(0,0)[lt]{\begin{minipage}{0.141806\unitlength}\raggedright flip-flops\end{minipage}}}%
    \put(0.8253152,0.17970507){\color[rgb]{0,0,0}\makebox(0,0)[lt]{\begin{minipage}{0.204985\unitlength}\raggedright $\mathit{ff_{k_3}}$\end{minipage}}}%
    \put(0,0){\includegraphics[width=\unitlength,page=6]{shift_arrival_gate.pdf}}%
  \end{picture}%
\endgroup%

\caption{Removal of a retimed flip-flop.}
\label{fig:shift_arrival}
}
\end{figure}

\textbf{Case 3:}
The net from gate $g_i$ to $g_j$ does not have a retimed flip-flop, $w_r(e_{g_i,g_j})= 0$. In 
this case, the data at the output of $g_i$ passes through $g_j$ directly.

When removing flip-flops combined with retiming,
each of the cases above can happen.
We let the solver determine which case actually happens during %
wave-pipelining construction. 
After this construction, no flip-flop should 
appear 
on %
the relevant paths. %
Consequently, if there is a retimed flip-flop on a net along a relevant path $w_r(e_{g_i,g_j})=1$, this flip-flop should be removed, 
so that $y_{e_{g_i,g_j}}=1$.  
Accordingly,   
the following constraints should be met 
\begin{equation}\label{eq:net_0}
w_r(e_{g_i,g_j})=y_{e_{g_i,g_j}}, \quad \forall e_{g_i,g_j} \in E \text{ on relevant paths}.
\end{equation}

We thus formulate the wave-pipelining construction problem as follows 
\begin{alignat}{2}\label{eq:first_formulation}
\mathrlap{\text{Minimize}}& \quad\quad\quad\quad\quad\quad \alpha \sum_{g \in G} \xi_{g}-\hlreview{\beta\sum_{g \in G}\sum^{n_g}_{i=1} d^i_g}\\
&+\gamma \sum_{g \in G} r(g)(\#input \, of \, g-\#output \,of\, g) \nonumber\\
&\rlap{\text{Subject to}} \nonumber \\ 
&\text{(\ref{eq:remove_ff}) and gray region constraints} && \label{eq:seonc} \\ %
&\text{Case 1 constraints, if $w_r(e_{g_i,g_j}) + y_{e_{g_i,g_j}}=1$} &&  \label{eq:first}\\
&\text{Case 2 constraints, if $w_r(e_{g_i,g_j}) + y_{e_{g_i,g_j}}=2$}  && \label{eq:sce} \\ 
&\text{Case 3 constraints,  if $w_r(e_{g_i,g_j}) + y_{e_{g_i,g_j}}=0$} && \label{eq:if_con}
\end{alignat}
where $\xi_{g}$ is introduced to enlarge the delay of wave-pipelining paths, which
can be
implemented by lengthening interconnects. 
\hlreview{$d^i_g$ is the delay from the $i$th input pin to the output pin of gate $g$. 
$n_g$ is the number of input pins for $g$.  }
$r(g)(\#input \, of \, g-\#output \,of\, g)$ represents the increased number of 
retimed flip-flops. 
$\alpha$, $\beta$ and $\gamma$ are constants with
$\alpha \ge \gamma \ge \beta$ to 
prevent the exposure of wave-pipelining resulting from lengthened interconnects and suppress area overhead by more 
retimed flip-flops and gate sizing. 
The conditional constraints (\ref{eq:if_con})--(\ref{eq:first}) 
can be converted into linear constraints as described in \cite{gurobi}. 

Since only one flip-flop is used to construct wave-pipelining at a time and  
other flip-flops are kept in the circuit as shown in \figname~\ref{fig:retiming1}(b), 
the part of a circuit around this flip-flop for  
wave-pipelining construction is not large.  
Therefore, we solve (\ref{eq:first_formulation})-(\ref{eq:first}) directly with an ILP solver to construct wave-pipelining. %

\subsection{ Wave-pipelining construction with duplication combined with retiming}\label{sec:wave_cons_duplication}

\begin{figure}[t]
{\figurefontsize
\centering
\begingroup%
  \makeatletter%
  \providecommand\color[2][]{%
    \errmessage{(Inkscape) Color is used for the text in Inkscape, but the package 'color.sty' is not loaded}%
    \renewcommand\color[2][]{}%
  }%
  \providecommand\transparent[1]{%
    \errmessage{(Inkscape) Transparency is used (non-zero) for the text in Inkscape, but the package 'transparent.sty' is not loaded}%
    \renewcommand\transparent[1]{}%
  }%
  \providecommand\rotatebox[2]{#2}%
  \ifx\svgwidth\undefined%
    \setlength{\unitlength}{227.80915205bp}%
    \ifx\svgscale\undefined%
      \relax%
    \else%
      \setlength{\unitlength}{\unitlength * \real{\svgscale}}%
    \fi%
  \else%
    \setlength{\unitlength}{\svgwidth}%
  \fi%
  \global\let\svgwidth\undefined%
  \global\let\svgscale\undefined%
  \makeatother%
  \begin{picture}(1,0.85939475)%
    \put(0,0){\includegraphics[width=\unitlength,page=1]{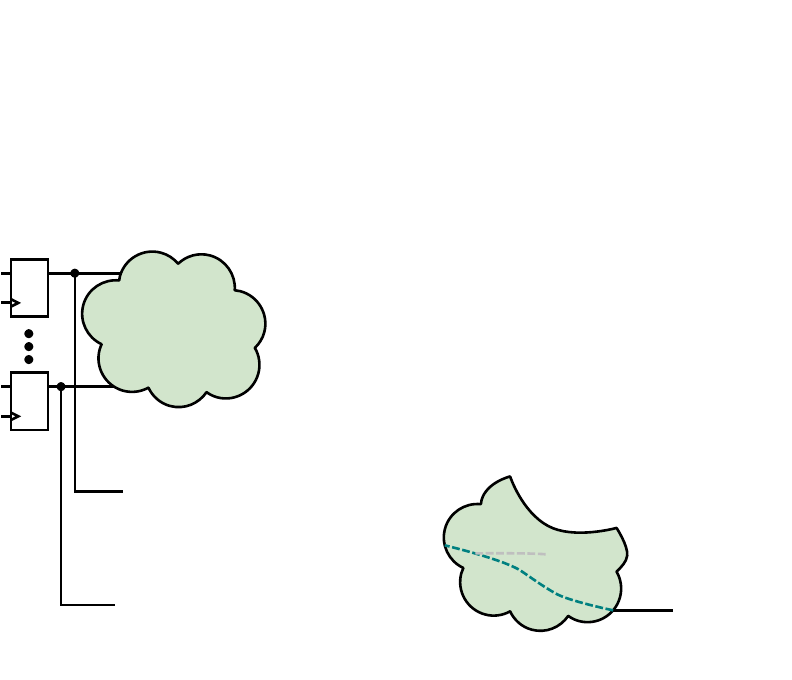}}%
    \put(0.65686365,0.1142102){\color[rgb]{0.82352941,0.89803922,0.8}\makebox(0,0)[b]{\smash{T}}}%
    \put(0,0){\includegraphics[width=\unitlength,page=2]{wp_cons.pdf}}%
    \put(0.87463205,0.12305234){\color[rgb]{0,0,0}\makebox(0,0)[b]{\smash{WP}}}%
    \put(0,0){\includegraphics[width=\unitlength,page=3]{wp_cons.pdf}}%
    \put(0.22226751,0.03614122){\color[rgb]{0,0,0}\makebox(0,0)[b]{\smash{duplicated}}}%
    \put(0.68756393,0.02955676){\color[rgb]{0,0,0}\makebox(0,0)[b]{\smash{duplicated}}}%
    \put(0,0){\includegraphics[width=\unitlength,page=4]{wp_cons.pdf}}%
    \put(0.40704032,0.07372034){\color[rgb]{0,0,0}\makebox(0,0)[b]{\smash{sized}}}%
    \put(0,0){\includegraphics[width=\unitlength,page=5]{wp_cons.pdf}}%
    \put(0.10519722,0.55628042){\color[rgb]{0,0,0}\makebox(0,0)[lb]{\smash{arrival times: $\overline{s}^c_{g_i}$, $\underline{s}^c_{g_i}$ }}}%
    \put(0.10519722,0.28224291){\color[rgb]{0,0,0}\makebox(0,0)[lb]{\smash{arrival times: $\overline{s}_{g_i}$, $\underline{s}_{g_i}$ }}}%
    \put(0.8935491,0.22867305){\color[rgb]{0,0,0}\makebox(0,0)[b]{\smash{maximum delay }}}%
    \put(0,0){\includegraphics[width=\unitlength,page=6]{wp_cons.pdf}}%
    \put(0.89400464,0.19508996){\color[rgb]{0,0,0}\makebox(0,0)[b]{\smash{of WP paths}}}%
    \put(0.72975878,-0.1588221){\color[rgb]{0,0,0}\makebox(0,0)[lt]{\begin{minipage}{0.38644101\unitlength}\raggedright \end{minipage}}}%
    \put(0.40617265,-0.10295115){\color[rgb]{0,0,0}\makebox(0,0)[lt]{\begin{minipage}{1.13138759\unitlength}\raggedright \end{minipage}}}%
    \put(0,0){\includegraphics[width=\unitlength,page=7]{wp_cons.pdf}}%
    \put(0.37930041,0.4671819){\color[rgb]{0,0,0}\makebox(0,0)[b]{\smash{$\mathit{ff_j}$}}}%
    \put(0.37930041,0.37109809){\color[rgb]{0,0,0}\makebox(0,0)[b]{\smash{$\mathit{ff_k}$}}}%
    \put(0,0){\includegraphics[width=\unitlength,page=8]{wp_cons.pdf}}%
    \put(0.68702469,0.43779643){\color[rgb]{0.82352941,0.89803922,0.8}\makebox(0,0)[b]{\smash{F}}}%
    \put(0.64674127,0.46650757){\color[rgb]{0.82352941,0.89803922,0.8}\makebox(0,0)[b]{\smash{T}}}%
    \put(0,0){\includegraphics[width=\unitlength,page=9]{wp_cons.pdf}}%
    \put(0.455269,0.60898438){\color[rgb]{0,0,0}\makebox(0,0)[b]{\smash{(a)}}}%
    \put(0.14839277,0.79839631){\color[rgb]{0,0,0}\makebox(0,0)[lt]{\begin{minipage}{0.33672249\unitlength}\raggedright relevant \end{minipage}}}%
    \put(0.37760924,0.78922645){\color[rgb]{0,0,0}\makebox(0,0)[b]{\smash{$\mathit{ff_j}$}}}%
    \put(0.37690689,0.69314254){\color[rgb]{0,0,0}\makebox(0,0)[b]{\smash{$\mathit{ff_k}$}}}%
    \put(0.66619319,0.79839631){\color[rgb]{0,0,0}\makebox(0,0)[lt]{\begin{minipage}{0.33672249\unitlength}\raggedright relevant \end{minipage}}}%
    \put(0,0){\includegraphics[width=\unitlength,page=10]{wp_cons.pdf}}%
    \put(0.19665643,0.14614035){\color[rgb]{0,0,0}\makebox(0,0)[lt]{\begin{minipage}{0.04190323\unitlength}\raggedright \end{minipage}}}%
    \put(0.15184326,0.19037154){\color[rgb]{0,0,0}\makebox(0,0)[lt]{\begin{minipage}{0.04888712\unitlength}\raggedright \end{minipage}}}%
    \put(0.24881865,0.12791616){\color[rgb]{0,0,0}\makebox(0,0)[lt]{\begin{minipage}{0.15364523\unitlength}\raggedright $g_j$\end{minipage}}}%
    \put(0.16264328,0.12791616){\color[rgb]{0,0,0}\makebox(0,0)[lt]{\begin{minipage}{0.15364523\unitlength}\raggedright $g_i$\end{minipage}}}%
    \put(0,0){\includegraphics[width=\unitlength,page=11]{wp_cons.pdf}}%
    \put(0.45610813,0.00524452){\color[rgb]{0,0,0}\makebox(0,0)[b]{\smash{(b)}}}%
    \put(0.61848803,0.34175738){\color[rgb]{0,0,0}\makebox(0,0)[lt]{\begin{minipage}{0.23279573\unitlength}\raggedright original\end{minipage}}}%
    \put(0.15319162,0.34175738){\color[rgb]{0,0,0}\makebox(0,0)[lt]{\begin{minipage}{0.23279573\unitlength}\raggedright original\end{minipage}}}%
  \end{picture}%
\endgroup%

\caption{Wave-pipelining construction with logic duplication and gate sizing combined with retiming. (a) $\mathit{ff_i}$ is moved to the left of the OR gate after retiming. (b) Logic duplication and gate sizing.}
\label{fig:wp_cons}
}
\end{figure}

The wave-pipelining construction by applying the removal of 
flip-flops combined with retiming described above might not be achieved successfully for a flip-flop $\mathit{ff_i}$, 
due to the circuit structure and the restriction on the area overhead incurred by lengthening 
interconnects, so that the ILP formulation (\ref{eq:first_formulation})--(\ref{eq:first}) 
may return no solution. %
To solve this problem, retiming is first applied to block irrelevant paths as much as possible 
and the circuit is duplicated in part to bypass such paths further to facilitate wave-pipelining construction. 
To implement the first step, 
the variable $y_{e}$, which indicates whether the retimed flip-flop on a net  
is removed, should be set to 0 for all nets in a circuit to guarantee that 
all paths are still single-period. %
The original flip-flop $\mathit{ff_i}$ should be moved to its left side 
to block irrelevant paths  
as much as possible with the formulation %
(\ref{eq:first_formulation})-(\ref{eq:first}), as shown in \figname~\ref{fig:wp_cons}(a). 
This leftward movement of flip-flops leads to fewer logic gates to be duplicated, as shown in \figname~\ref{fig:wp_cons}(b).

\begin{table*}
\renewcommand{\arraystretch}{1.1}
\renewcommand{\tabcolsep}{4.18pt}
\centering
\renewcommand{\arraystretch}{1.05}
\renewcommand{\tabcolsep}{6.8pt}
\caption{Results of Constructing Wave-pipelining Paths}
\label{tb_test}
\begin{tabular}{lcc c ccccccccc cc c c} \hlinewd{0.7pt}

\multicolumn{3}{c}{Circuit} &

\multicolumn{1}{c}{} &

\multicolumn{3}{c}{WP True Construction} &

\multicolumn{1}{c}{} &

\multicolumn{3}{c}{WP False Construction} &

\multicolumn{1}{c}{} &

\multicolumn{3}{c}{Cost} &

\multicolumn{1}{c}{} &

\multicolumn{1}{c}{Runtime}  \\

\cline {1-3} \cline{5-7} \cline{9-11} \cline{13-15} \cline {17-17}

\multicolumn{1}{c}{} &
\multicolumn{1}{c}{$n_s$} &
\multicolumn{1}{c}{$n_g$} &

\multicolumn{1}{c}{} &

\multicolumn{1}{c}{$n_{wpt}$} & 
\multicolumn{1}{c}{$n_{t}$} &
\multicolumn{1}{c}{$n^\prime_t$} & 
\multicolumn{1}{c}{} &
\multicolumn{1}{c}{$n_{wpf}$} &
\multicolumn{1}{c}{$n_{f}$} &
\multicolumn{1}{c}{$n^\prime_f$} &
\multicolumn{1}{c}{} & 
\multicolumn{1}{c}{$n_{p}$} &
\multicolumn{1}{c}{$n_{d}$} &
\multicolumn{1}{c}{$n_{r}$} &

\multicolumn{1}{c}{} &

\multicolumn{1}{c}{$t_r(s)$} \\

\hlinewd{0.6pt}

s35932 &1728 &16065  &  &579 &80213  &80792 & &715 &130087 &130802 & & 23.4 &35  &2  &&481.8 \\ 
s38584 &1452 &19253  &  &420 &202647  &203067 & &2486 &722378 &724864 & &9.7 &63 &4  &&736.2 \\
s38417 &1636 &22179  &  &8369 &637091  &645460 & &9837 &594078 &603915 & &18.9 &105  &0  &&643.8 \\
s15850 &534  &9772   &  &1932 &1693510  &1695442 & &606 &20999926 &21000532 & & 98.6 &0   &4 &&699.0 \\
s13207 &669  &7951   &  &7830 &294531   &302361& &13410 &728262 &741672 & & 12.5 &74  &4 &&223.4 \\
s9234  &228  &5597   &  &282 &5710   &5992 & &1349 &154825 &156174 & & 27.2 &62  &2  &&362.3 \\
s5378  &179  &2779   &  &844 &8465  &9309 & &706 &467 &1173 & & 49.3 &73   &0  &&114.9 \\
s4863  &104  &2342   &  &300 &359663  &359963 & &0 &70432413 &70432413 & & 38.9    &26  &5 &&3564.7\\
s1423  &74   &657    &  &278 &7997 &8275 & &272 &5698 &5970 & & 45.7 &91 &2 &&19.3 \\
s1238  &18   &508    &  &4 &381   &385 & &2  &849 &851    & & 19.9 &14     &5 &&0.8   \\

\hlinewd{0.7pt}
\end{tabular}
\end{table*}

In the second step, after retiming, 
we duplicate the logic in the circuit and size the gates and lengthen interconnects 
so that the delays of all wave-pipelining paths meet timing constraints 
(\ref{eq:short_path})--(\ref{eq:gray}) as illustrated in 
\figname~\ref{fig:wp_cons}(b). In the duplicated circuit on the right of
retimed flip-flops, we only keep the flip-flops at which wave-pipelining
paths terminate. The other flip-flops stay in the original circuit. Afterwards, we
delete the logic gates backwards to remove those gates that do not drive any
flip-flop %
to reduce resource usage. 
When duplicating the logic on the left of $\mathit{ff_j}$, however, we need to keep
all the logic gates 
to maintain the correct function of the circuit.

In the duplicated logic in \figname~\ref{fig:wp_cons}(b), we do not duplicate 
flip-flops. Therefore, all combinational paths in the
duplicated logic are wave-pipelining paths and
their delays should meet the timing constraints (\ref{eq:short_path})--(\ref{eq:gray}).
To meet these constraints, we size the gates and lengthen interconnects 
in the duplicated logic with an
ILP formulation.
For example, in \figname~\ref{fig:wp_cons}(b), we assume the latest and the earliest arrival 
times at the output of the inverter in the duplicated circuit as $\overline{s}_{g_i}$ and $\underline{s}_{g_i}$. 
Similarly, we assume the latest and the earliest arrival times at the output of 
the AND gate in the duplicated circuit as as $\overline{s}_{g_j}$ and $\underline{s}_{g_j}$. 
\hlreview{Furthermore, the delay from an input pin to the output pin of the AND gate along the path $p$ 
traveling through the inverter 
is written as $d^p_{g_j}$.}
With these definitions, the arrival times 
between the inverter and the AND gate 
can be written as
\hlreview{
\begin{align}
\label{eq:late_arr}
\overline{s}_{g_j} &\ge \overline{s}_{g_i}+ \xi_{g_j}+d^p_{g_j} \\
\label{eq:early_arr}
\underline{s}_{g_j} &\le \underline{s}_{g_i}+ \xi_{g_j}+d^p_{g_j}.
\end{align}
}
To reduce the number of duplicated gates, we try to connect the output pins of logic gates in
the duplicated logic to the original gates as much as possible, as
illustrated in \figname~\ref{fig:wp_cons}(b). 
In the original logic, the latest and the earliest arrival times are
constants. 
Assume that the arrival times at the output of the inverter in the original circuit 
 are $\overline{s}^c_{g_i}$ and $\underline{s}^c_{g_i}$, and
a 0-1 variable $p_i$ indicates whether the output pin in the
duplicated logic should be driven by the original logic.
We can then %
extend the constraints (\ref{eq:late_arr})--(\ref{eq:early_arr}) as
\hlreview{
\begin{align}
\label{eq:late_arr_s1}
\overline{s}_{g_j} &\ge \overline{s}_{g_i}+d^p_{g_j}+\xi_{g_j}-p_i{M} \\
\label{eq:late_arr_s2}
\overline{s}_{g_j} &\ge \overline{s}^c_{g_i}+d^p_{g_j}+\xi_{g_j}-(1-p_i){M} \\
\label{eq:early_arr_e1}
\underline{s}_{g_j} &\le \underline{s}_{g_i}+d^p_{g_j}+\xi_{g_j}+p_i{M}\\
\label{eq:early_arr_e2}
\underline{s}_{g_j} &\le \underline{s}^c_{g_i}+d^p_{g_j}+\xi_{g_j}+(1-p_i){M}
\end{align}
}
where ${M}$ is a very large positive constant used to transform the
conditional constraints to linear constraints \cite{chen2011applied}. In
either case when an output pin in the duplicated logic is connected or disconnected with the original
logic, only two constraints in (\ref{eq:late_arr_s1})--(\ref{eq:early_arr_e2})
are valid.

In the description above, %
we allow a path delay to be extended with lengthening interconnects. 
However, we try to keep the delay incurred by interconnects as small as possible.
In addition, we try to reduce the overall area overhead when implementing wave-pipelining. 
Therefore, 
we formulate the construction problem as
\begin{align} 
\text{minimize} & \quad \alpha \sum_{g\in G} \xi_{g} -\hlreview{\beta\sum_{g \in G}\sum^{n_g}_{i=1} d^i_g}-\gamma\sum_{i\in I}p_{i} \label{eq:minobj}\\
\text{subject to} & \quad
 \text{(\ref{eq:f_short_path})--(\ref{eq:f_long_path}) \text{,} (\ref{eq:gray_arr_c})} \text{ and (\ref{eq:late_arr_s1})--(\ref{eq:early_arr_e2})}\label{eq:conds}
\end{align} 
where $\alpha \ge \gamma \ge \beta$ to prevent the exposure of wave-pipelining 
resulting from lengthened interconnects and duplicated gates and suppress area overhead by gate sizing. %
In this setting, the effectiveness of camouflaging is more important than incurred area overhead.
$I$ is the index set of all output pins.
After the ILP problem above is solved, the gates that
do not drive any other gates in the duplicated logic are removed to
reduce resource usage. %
With this extra step, wave-pipelining can also be constructed even if the optimization problem in (\ref{eq:first_formulation})-(\ref{eq:first}) 
returns no solution.

In the ILP formulation
  (\ref{eq:first_formulation})--(\ref{eq:if_con}) and
  (\ref{eq:minobj})--(\ref{eq:conds}), we assume constant input slews
  and output loads for gate delays for simplification. 
 \hlreview{ 
However, we do not know which combination of input slews and
output loads should be adopted for the delay reference of a gate in the ILP formulation.
For simplification,
the input slew and output load values in the middle of their 
corresponding ranges in the lookup table are used as typical values 
to select the delay for wave-pipelining construction.
After the wave-pipelining paths are determined, we 
verify the timing constraints of these paths using real lookup tables 
indexed by slew and load values. For each input pin to the output pin of a
gate, the delay and output slew are set according to real input slew and output
load using the corresponding lookup tables.
Due to the assumption of constant delays in the ILP formulation, 
the real delays of wave-pipelining paths 
may not meet the timing constraints.   
In case of timing violation, %
we first size the logic gates
iteratively to modify the path delay. If this is still not 
sufficient to
solve the
timing violation, buffers are then added or removed to meet
wave-pipelining constraints.
If the
timing constraints are still not met with 
gate sizing and buffer insertion described above, we
switch to another flip-flop to construct wave-pipelining paths.  
However, the construction process is not guaranteed to converge. 
}

\section{Experimental Results}\label{sec:results}

TimingCamouflage+ was implemented in C++ and tested
using a \SI[mode=text]{3.20}{\GHz} CPU. %
We demonstrate the results using circuits
from the ISCAS89 benchmark set.
The number of flip-flops and the number of logic gates are shown in the  
columns $\boldsymbol{n_s}$ and $\boldsymbol{n_g}$ in Table~\ref{tb_test}, respectively.
The benchmark circuits were sized using a \SI[mode=text]{45}{\nm} library.
We set the timing margin $\delta$ to 0.15 to tolerate PVT variations
and the inaccuracy factor $\tau$ of delay estimation of attackers to 0.2. 
To simplify the delay models,
input slews and output loads are set to constant values.
However, TimingCamouflage+
is independent of delay characterization and can work with any delay model. 
We used Gurobi \cite{gurobi} to solve the optimization problems. %

The results of wave-pipelining path construction are shown in Table~\ref{tb_test}. 
The column $\boldsymbol{n_{wpt}}$ shows the numbers of wave-pipelining true paths
whose delays are in the gray region. These paths are used to guarantee that attackers 
must perform testing to distinguish wave-pipelining true paths 
from single-period clocking true paths. 
Without these paths, attackers can assume all testable paths are single-period and avoid 
the expensive test procedure.    
The
column $\boldsymbol{n_t}$ shows the numbers of single-period clocking true paths 
whose delays meet the gray region requirement (\ref{eq:gray}).
When attackers try to detect the
locations of wave-pipelining paths, these true paths need to be tested
to determine whether their delays are actually larger or smaller than $T$. 
The column 
$\boldsymbol{n^\prime_{t}}=\boldsymbol{n_{wpt}}+\boldsymbol{n_t}$,  
is the total number of suspicious true paths that are 
required to be tested.    

The column $\boldsymbol{n_{wpf}}$ shows the 
numbers of wave-pipelining false paths whose delays are in the gray region. 
In the experiments, we set the target numbers of
wave-pipelining true and false paths both to 100 and 
the threshold distance $dis_t$ between the flip-flops in the first line of \figname~\ref{fig:work_flow} to construct wave-pipelining 
true and false paths to 10 times of the minimum distance between all pairs of 
flip-flops.  
We executed the construction of wave-pipelining true
and false paths shown in \figname~\ref{fig:work_flow} repeatedly %
using the method described in Section~\ref{sec:wave_cons_removal} and Section~\ref{sec:wave_cons_duplication}. 
When we constructed wave-pipelining false paths,  
we also found wave-pipelining true paths in
the circuit and vice versa. 
Consequently, the numbers of 
these paths shown in the columns $\boldsymbol{n_{wpt}}$ and $\boldsymbol{n_{wpf}}$ are larger than 100
for many test cases except s4863 and s1238. In s4863 there is no
wave-pipelining false path and in s1238 the number of wave-pipelining paths
is very
small due to the limited sizes of these two circuits. In all large
test cases, however, wave-pipelining paths have been constructed successfully.
In practice, as the circuit size increases, 
more path candidates become available for the wave-pipelining false path construction,
so that wave-pipelining false paths can always be constructed successfully. 

The column $\boldsymbol{n_f}$ shows the number of suspicious single-period clocking false paths. 
Since their delays meet the gray region requirement (\ref{eq:gray}), 
these paths are suspicious wave-pipelining false paths for attackers. 
The total suspicious wave-pipelining false paths are shown in 
$\boldsymbol{n^\prime_{f}}=\boldsymbol{n_{wpf}}+\boldsymbol{n_f}$. 
To determine whether a false path is wave-pipelining or not from a huge number of suspicious paths, 
much effort is required as explained in Section~\ref{sec:counter_measures}, 
since these false paths cannot be triggered for delay test.

The column $\boldsymbol{n_p}$ shows the equivalent number of inserted 
delays in the unit of the delay of a buffer 
by lengthening interconnects to enlarge wave-pipelining path delays.
Since the number of inserted delays does not increase with respect to circuit
size, the area cost for constructing wave-pipelining paths is negligible in
relatively large circuits.

In the experiments, we constrained the inserted delay $\xi_{g}$ incurred by lengthening 
interconnects to be no larger than
the delay of three buffers for each net. 
With this constraint, the removal of flip-flops combined with retiming in Section~\ref{sec:wave_cons_removal} 
can only be achieved in s15850 to construct wave-pipelining paths. %
In the remaining circuits,
this method cannot construct wave-pipelining successfully, 
since a few short paths still require much delay insertion to meet the timing constraint of wave-pipelining. However, 
the number of such paths is very small, confirmed by the total delays inserted as shown in column $\boldsymbol{n_p}$. 
In this
case, %
circuit duplication combined with retiming is applied as described in Section~\ref{sec:wave_cons_duplication}. 
The column $\boldsymbol{n_d}$ in Table~\ref{tb_test} shows the number of logic gates duplicated 
to construct wave-pipelining. Since we only inserted wave-pipelining
paths at a limited number of locations, %
generally the number of duplicated gates does not increase with respect to circuit size. 

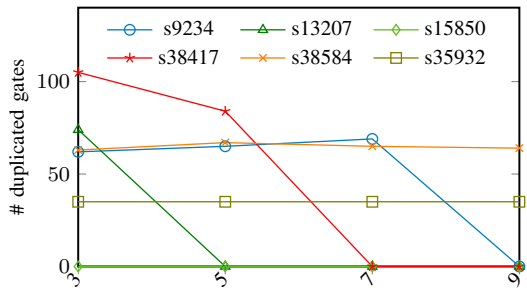
\begin{figure}[t]
{
\figurefontsize
\centering
\pgfplotsset{compat=1.3,
    /pgfplots/ybar legend/.append style={ 
        /pgfplots/legend image code/.code={%
           \draw[##1,/tikz/.cd,yshift=-0.25em]
           (0cm,0cm) rectangle (5pt,0.8em);
        },
    }
}

\begin{tikzpicture}

\pgfplotstableread[row sep=\\] {
circuit s9234 s13207 s15850 s38417 s38584 s35932 \\
1        62    74      0     105     63    35\\
2        65    0       0     84      67    35\\
3        69    0       0     0       65    35\\
4        0     0       0     0       64    35\\
}\loadedtable

\begin{axis}[
xticklabels={3,5,7,9},
xtick={1,...,4},
xmin=1, xmax=4.0,
x=1.955cm, y=0.0246cm, 
x tick label style={rotate=45, xshift=0pt,yshift=0pt,anchor=east, 
inner sep=0}, 
xticklabel pos=left, xtick align=outside, xtick pos=left,
ymin=0, ymax=140, 
ylabel={\# duplicated gates}, 
ylabel style={inner sep=0}, 
ylabel shift=0pt, ytickmin=0,ytickmax=140, 
legend columns=3, 
legend style={
at={(0.5,0.86)}, anchor=center, 
/tikz/every even column/.append style={column sep=0.2cm},
draw=none, 
},
line width=0.75pt,
major tick length=3pt,
]  \addplot[sharp plot, line width=0.5pt, cyan!60!blue, mark=o, fill=none] table[x=circuit,y=s9234] {\loadedtable};
   \addplot[sharp plot, line width=0.5pt, green!50!black, mark=triangle, fill=none] table[x=circuit,y=s13207] {\loadedtable};
   \addplot[sharp plot, line width=0.5pt, green!50!brown, mark=diamond, fill=none] table[x=circuit,y=s15850] {\loadedtable};
   \addplot[sharp plot, line width=0.5pt, red!50!red, mark=star, fill=none] table[x=circuit,y=s38417] {\loadedtable};
   \addplot[sharp plot, line width=0.5pt, red!50!yellow, mark=x, fill=none] table[x=circuit,y=s38584] {\loadedtable};
   \addplot[sharp plot, line width=0.5pt, red!50!green, mark=square, fill=none] table[x=circuit,y=s35932] {\loadedtable};
   \legend{s9234,s13207,s15850,s38417,s38584,s35932}
\end{axis}
\end{tikzpicture}

\caption{The number of duplicated gates over the increase of interconnect delays (in unit of buffer delays).}
\label{fig:buffer_d}
}
\end{figure}

To construct wave-pipelining paths, retiming %
might cause an increase 
of the number of flip-flops, shown in column $\boldsymbol{n_r}$. 
This increase %
and the corresponding area overhead are still negligible.     
The last column $\boldsymbol{t_r}$ in Table~\ref{tb_test} shows the runtime of TimingCamouflage+, 
which is acceptable because wave-pipelining construction is a one-time
effort. 
 
We also applied TimingCamouflage+ in two practical circuits, vga\_lcd and pci\_bridge32, from TAU 2013 variation-aware timing analysis contest. 
In both circuits, wave-pipelining false and true paths can be constructed successfully. 
Specifically, 
in vga\_lcd, 169 wave-pipelining  
true paths and 138 wave-pipelining false paths were 
constructed with 25 inserted buffers, 60 duplicated gates and 
5 more retimed flip-flops.  
In pci\_bridge32, 206 
wave-pipelining true paths and 1162 wave-pipelining false 
paths were constructed with 18 inserted buffers, 43 duplicated 
gates and 2 more retimed flip-flops.

When constructing wave-pipelining paths by lengthening interconnects, 
we constrained the interconnect delay $\xi_g$ inserted before a gate $g$. 
\figname~\ref{fig:buffer_d} 
shows the changes of the numbers of duplicated gates when %
the interconnect delays are relaxed. 
It is clear that with the increase of delays inserted before a gate $g$, 
the number of duplicated gates required to construct wave-pipelining paths is decreased in three circuits. 
When the allowable number of delay units  
reaches 9, no circuit duplication is required for all these circuits. 
In s15850, 
duplicated gates are not even required to construct the wave-pipelining paths when 
the allowable number of delay units is larger than or equal to 3, due to the efficiency of 
the removal of flip-flops combined with retiming in Section~\ref{sec:wave_cons_removal}. 
For s38584 and s35932, the removal of flip-flops 
combined with retiming cannot be achieved successfully due to 
circuit structures. Therefore, the numbers of duplicated gates are not 
reduced.

\begin{figure}[t]
{
\figurefontsize
\centering
\pgfplotsset{compat=1.3,
    /pgfplots/ybar legend/.append style={ 
        /pgfplots/legend image code/.code={%
           \draw[##1,/tikz/.cd,yshift=-0.25em]
           (0cm,0cm) rectangle (7pt,0.8em);
        },
    }
}

\begin{tikzpicture}

\pgfplotstableread[row sep=\\] {
circuit  Originally             Reduced \\
1        2.08                      0.35\\
2        6.37                      0.63\\
3        6.15                     1.05\\
4        0.64                       0\\
5        9.28                     0.74\\
6        3.74                     0.62\\
7        2.07                     0.73\\
8        7.44                     0.26\\
9        1.61                     0.91\\
10       0.57                      0.14\\
}\loadedtable

\begin{axis}[
xticklabels={s1238,s1423,s4863,s5378,s9234,s13207,s15850,s38417,s38584,s35932},
xtick={1,...,10},
xmin=0, xmax=10.4,
x=0.63cm, y=0.1973cm, 
x tick label style={rotate=45, xshift=0pt,yshift=0pt,anchor=east, 
inner sep=0}, 
xticklabel pos=left, xtick align=outside, xtick pos=left,
ymin=0, ymax=13,
ylabel={Num.of gates ($\times 10^{2}$)}, 
ylabel style={inner sep=0}, 
ylabel shift=0pt, ytickmin=0,ytickmax=13, 
legend columns=3, 
legend style={
at={(0.45,0.880)}, anchor=center, 
/tikz/every even column/.append style={column sep=0.2cm},
draw=none, 
},
line width=0.75pt,
ybar=0pt, 
bar width=6pt,
axis on top=true,
major tick length=3pt,
]  
\addplot[ybar, line width=0.35pt, red, fill=red!30!white] table[x=circuit,y=Originally] {\loadedtable};
\addplot[ybar, line width=0.35pt, blue, fill=blue!50!white] table[x=circuit,y=Reduced] {\loadedtable};
\legend{Before red., After red.}

\end{axis}
\end{tikzpicture}

\caption{Comparison of gate numbers before/after reduction.}
\label{fig:gate_num}
}
\end{figure}
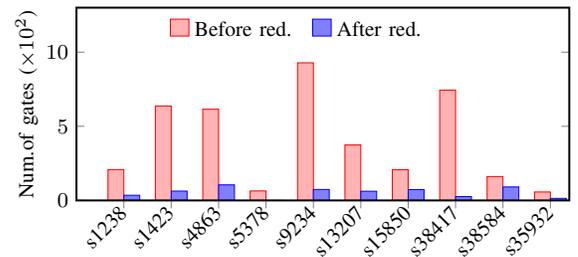

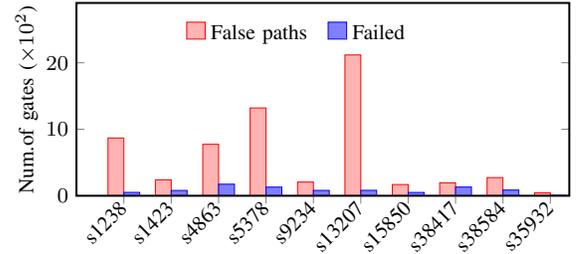
\begin{figure}[t]
{
\figurefontsize
\centering
\pgfplotsset{compat=1.3,
    /pgfplots/ybar legend/.append style={ 
        /pgfplots/legend image code/.code={%
           \draw[##1,/tikz/.cd,yshift=-0.25em]
           (0cm,0cm) rectangle (7pt,0.8em);
        },
    }
}

\begin{tikzpicture}

\pgfplotstableread[row sep=\\] {
circuit  False                 Failed \\
1        8.69                     0.51\\
2        2.40                     0.80\\
3        7.74                     1.74\\
4        13.21                    1.30\\
5        2.06                     0.81\\
6        21.18                    0.82\\
7        1.69                     0.50\\
8        1.95                     1.31\\
9        2.73                     0.86\\
10       0.43                      0.07\\
}\loadedtable

\begin{axis}[
xticklabels={s1238,s1423,s4863,s5378,s9234,s13207,s15850,s38417,s38584,s35932},
xtick={1,...,10},
xmin=0, xmax=10.4,
x=0.63cm, y=0.08845cm, 
x tick label style={rotate=45, xshift=0pt,yshift=0pt,anchor=east, 
inner sep=0}, 
xticklabel pos=left, xtick align=outside, xtick pos=left,
ymin=0, ymax=29,
ylabel={Num.of gates ($\times 10^{2}$)}, 
ylabel style={inner sep=0}, 
ylabel shift=0pt, ytickmin=0,ytickmax=29, 
legend columns=3, 
legend style={
at={(0.45,0.84)}, anchor=center, 
/tikz/every even column/.append style={column sep=0.2cm},
draw=none, 
},
line width=0.75pt,
ybar=0pt, 
bar width=6pt,
axis on top=true,
major tick length=3pt,
]  
\addplot[ybar, line width=0.35pt, red, fill=red!30!white] table[x=circuit,y=False] {\loadedtable};
\addplot[ybar, line width=0.35pt, blue, fill=blue!50!white] table[x=circuit,y=Failed] {\loadedtable};
\legend{False paths, Failed}

\end{axis}
\end{tikzpicture}

\caption{Results of false path sizing attack.}
\label{fig:false_sizing}
}
\end{figure}
\begin{figure}[t]
{
\figurefontsize
\centering
\pgfplotsset{compat=1.3,
    /pgfplots/ybar legend/.append style={ 
        /pgfplots/legend image code/.code={%
           \draw[##1,/tikz/.cd,yshift=-0.25em]
           (0cm,0cm) rectangle (5pt,0.8em);
        },
    }
}

\begin{tikzpicture}

\pgfplotstableread [row sep=\\] {
circuit  Originally            Reduced \\
1        nan                   14\\
2        112                   91\\
3        nan                   26\\
4        95                    73\\
5        113                   62\\
6        107                  74\\
7        nan                   0\\
8        71                   105\\
9        69                   63\\
10       nan                   35\\
}\loadedtable

\begin{axis}[
xticklabels={s1238,s1423,s4863,s5378,s9234,s13207,s15850,s38417,s38584,s35932},
xtick={1,...,10},
xmin=1, xmax=10,
x=0.728cm, y=0.0171cm, 
x tick label style={rotate=45, xshift=0pt,yshift=0pt,anchor=east, 
inner sep=0}, 
xticklabel pos=left, xtick align=outside, xtick pos=left,
ymin=0, ymax=150, 
ylabel={\#duplicated gates}, 
ylabel style={inner sep=0}, 
ylabel shift=0pt, ytickmin=0,ytickmax=150, 
legend columns=3, 
legend style={
at={(0.5,0.89)}, anchor=center, 
/tikz/every even column/.append style={column sep=0.2cm},
draw=none, 
},
line width=0.75pt,
major tick length=3pt,
]  \addplot[unbounded coords=jump,sharp plot, line width=0.5pt, cyan!60!blue, mark=o, fill=none] table[x=circuit,y=Originally] {\loadedtable};
   \addplot[unbounded coords=jump,sharp plot, line width=0.5pt, green!50!black, mark=triangle, fill=none] table[x=circuit,y=Reduced] {\loadedtable};
   \legend{TimingCamouflage, TimingCamouflage+}
\end{axis}
\end{tikzpicture}

\caption{Comparison of the number of duplicated gates with TimingCamouflage in \cite{ZLYPS18}.}
\label{fig:gate_DATE}
}
\end{figure}
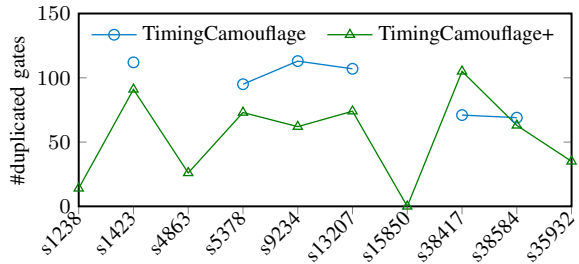

\begin{figure}[t]
{
\figurefontsize
\centering
\vspace{-0.6em}
\pgfplotsset{compat=1.3,
    /pgfplots/ybar legend/.append style={ 
        /pgfplots/legend image code/.code={%
           \draw[##1,/tikz/.cd,yshift=-0.25em]
           (0cm,0cm) rectangle (5pt,0.8em);
        },
    }
}

\begin{tikzpicture}

\pgfplotstableread[row sep=\\] {
circuit  Originally             Reduced \\
1        nan                     19.8\\
2        73                      43.5\\
3        nan                     38.9\\
4        66                      47.6\\
5        67                      24.4\\
6        57                      5.8\\
7        nan                     96.4\\
8        38                      13.0\\
9        21                       7.3\\
10       nan                     22.4\\
}\loadedtable
\begin{axis}[
xticklabels={s1238,s1423,s4863,s5378,s9234,s13207,s15850,s38417,s38584,s35932},
xtick={1,...,10},
xmin=1, xmax=10,
x=0.728cm, y=0.0171cm, 
x tick label style={rotate=45, xshift=0pt,yshift=0pt,anchor=east, 
inner sep=0}, 
xticklabel pos=left, xtick align=outside, xtick pos=left,
ymin=0, ymax=150, 
ylabel={\#inserted buffers}, 
ylabel style={inner sep=0}, 
ylabel shift=0pt, ytickmin=0,ytickmax=150, 
legend columns=3, 
legend style={
at={(0.5,0.87)}, anchor=center, 
/tikz/every even column/.append style={column sep=0.2cm},
draw=none, 
},
line width=0.75pt,
major tick length=3pt,
]  \addplot[unbounded coords=jump,sharp plot, line width=0.5pt, cyan!60!blue, mark=o, fill=none] table[x=circuit,y=Originally] {\loadedtable};
   \addplot[unbounded coords=jump,sharp plot, line width=0.5pt, green!50!black, mark=triangle, fill=none] table[x=circuit,y=Reduced] {\loadedtable};
   \legend{TimingCamouflage, TimingCamouflage+}
\end{axis}
\end{tikzpicture}

\caption{Comparison of the number of inserted delay units with TimingCamouflage in \cite{ZLYPS18}.}
\label{fig:buffer_DATE}
}
\end{figure}
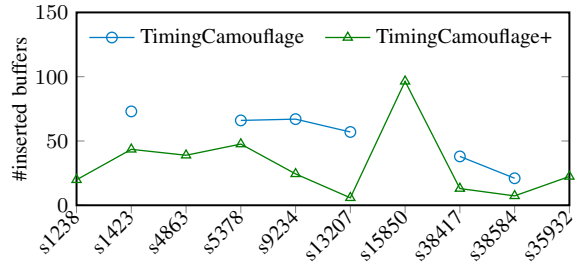

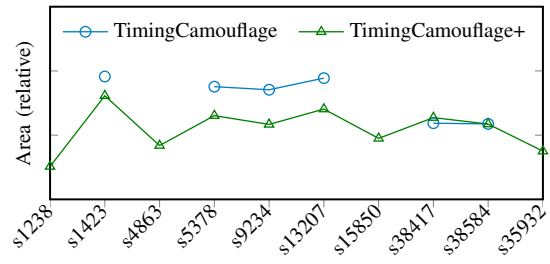
\begin{figure}[t]
{
\figurefontsize
\centering
\pgfplotsset{compat=1.3,
    /pgfplots/ybar legend/.append style={ 
        /pgfplots/legend image code/.code={%
           \draw[##1,/tikz/.cd,yshift=-0.25em]
           (0cm,0cm) rectangle (5pt,0.8em);
        },
    }
}

\begin{tikzpicture}

\pgfplotstableread [row sep=\\] {
circuit  Originally             Reduced \\
1        nan                     51.2\\
2        191.3                   161.4\\
3        nan                     83.8\\
4        175.5                   130.4\\
5        170.7                   116.7\\  
6        188.8                   140.6\\
7        nan                     95.0\\
8        118.6                   127.3\\
9        117.5                   117.3\\
10       nan                     75.1\\
}\loadedtable
\begin{axis}[
xticklabels={s1238,s1423,s4863,s5378,s9234,s13207,s15850,s38417,s38584,s35932},
xtick={1,...,10},
xmin=1, xmax=10,
x=0.728cm, y=0.00855cm, 
x tick label style={rotate=45, xshift=0pt,yshift=0pt,anchor=east, 
inner sep=0}, 
xticklabel pos=left, xtick align=outside, xtick pos=left,
ymin=0, ymax=300, 
ylabel={Area (relative)}, 
ylabel style={inner sep=0}, 
ylabel shift=0pt, ytickmin=0,ytickmax=300, 
yticklabels={,,},
legend columns=3, 
legend style={
at={(0.5,0.87)}, anchor=center, 
/tikz/every even column/.append style={column sep=0.2cm},
draw=none, 
},
line width=0.75pt,
major tick length=3pt,
]  \addplot[unbounded coords=jump,sharp plot, line width=0.5pt, cyan!60!blue, mark=o, fill=none] table[x=circuit,y=Originally] {\loadedtable};
   \addplot[unbounded coords=jump,sharp plot, line width=0.5pt, green!50!black, mark=triangle, fill=none] table[x=circuit,y=Reduced] {\loadedtable};
   \legend{TimingCamouflage, TimingCamouflage+}
\end{axis}
\end{tikzpicture}
 
\caption{Comparison of area overhead with TimingCamouflage in \cite{ZLYPS18}.}
\label{fig:area_DATE}
}
\end{figure}

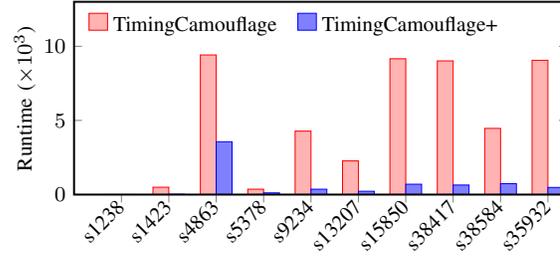
\begin{figure}[t]
{
\figurefontsize
\centering
\pgfplotsset{compat=1.3,
    /pgfplots/ybar legend/.append style={ 
        /pgfplots/legend image code/.code={%
           \draw[##1,/tikz/.cd,yshift=-0.25em]
           (0cm,0cm) rectangle (7pt,0.8em);
        },
    }
}

\begin{tikzpicture}

\pgfplotstableread[row sep=\\] {
circuit  Originally             Reduced \\
1        0.0012                    0.0008\\
2        0.4954                     0.0193\\
3        9.4150                    3.5647\\
4        0.3631                     0.1149\\
5        4.2796                    0.3623\\
6        2.2710                    0.2234\\ 
7        9.1607                    0.6990\\
8        9.0156                    0.6438\\
9        4.4629                    0.7362\\
10       9.0557                    0.4818\\
}\loadedtable

\begin{axis}[
xticklabels={s1238,s1423,s4863,s5378,s9234,s13207,s15850,s38417,s38584,s35932},
xtick={1,...,10},
xmin=0, xmax=10.4,
x=0.63cm, y=0.1973cm, 
x tick label style={rotate=45, xshift=0pt,yshift=0pt,anchor=east, 
inner sep=0}, 
xticklabel pos=left, xtick align=outside, xtick pos=left,
ymin=0, ymax=13.00,
ylabel={Runtime ($\times 10^{3}$)}, 
ylabel style={inner sep=0}, 
ylabel shift=0pt, ytickmin=0,ytickmax=13.00, 
legend columns=3, 
legend style={
at={(0.45,0.880)}, anchor=center, 
/tikz/every even column/.append style={column sep=0.2cm},
draw=none, 
},
line width=0.75pt,
ybar=0pt, 
bar width=6pt,
axis on top=true,
major tick length=3pt,
]  
\addplot[ybar, line width=0.35pt, red, fill=red!30!white] table[x=circuit,y=Originally] {\loadedtable};
\addplot[ybar, line width=0.35pt, blue, fill=blue!50!white] table[x=circuit,y=Reduced] {\loadedtable};
\legend{TimingCamouflage, TimingCamouflage+}

\end{axis}
\end{tikzpicture}

\caption{Comparison of runtime with TimingCamouflage in \cite{ZLYPS18}.}
\label{fig:runtime_DATE}
}
\end{figure}

In the wave-pipelining construction formulation (\ref{eq:minobj})--(\ref{eq:conds}), we
maximize the number of output pins that can be driven by the original circuit 
as illustrated in \figname~\ref{fig:wp_cons}(b). 
Consequently, the number of logic gates in the duplicated circuit can be reduced. 
\figname~\ref{fig:gate_num} compares the
numbers of gates in the originally duplicated circuit before the removed
flip-flop in \figname~\ref{fig:wp_cons}
and the number of gates after reduction by using the original circuit. 
In all test cases, the numbers of duplicated gates were reduced
significantly.

In the experiments, we also simulated the gate sizing attack
on the netlist as discussed in
Section~\ref{sec:counter_measures}. The basic idea was that all false paths
with delays falling in the gray region
were treated as wave-pipelining paths and their delays were sized to 
meet (\ref{eq:short_path})--(\ref{eq:long_path}).
The results of this simulated attack are shown in
\figname~\ref{fig:false_sizing}, where the first bar shows the number of false
paths we used to simulate the attack. 
The second bar shows the number of
false paths that were not sized successfully. %
Even with such a small number of false paths from the huge set of false paths in the original circuit, 
no
sizing attack succeeded. %

TimingCamouflage in \cite{ZLYPS18} uses the same timing concept to camouflage the netlists. However, 
it constructs wave-pipelining paths with duplicated gates without applying the retiming technique. 
Consequently, it requires 
many duplicated gates and interconnect delays to construct the wave-pipelining paths. 
To verify the improvement of incorporating retiming in 
TimingCamouflage+, TimingCamouflage in \cite{ZLYPS18} was implemented 
with the maximum allowable inserted delay units constrained the same as in TimingCamouflage+. 
The results are shown in \figname~\ref{fig:gate_DATE} and \figname~\ref{fig:buffer_DATE}. 
Since the maximum numbers of inserted delay units were constrained to be no larger than three, 
TimingCamouflage in \cite{ZLYPS18} fails to construct the wave-pipelining paths in s1238, 
s4863, s15850 and s35932, as shown in \figname~\ref{fig:gate_DATE}. 
In the other cases, the numbers of duplicated gates with TimingCamouflage+ are also smaller than those with TimingCamouflage 
in \cite{ZLYPS18}, except for s38417, 
in which the number of duplicated gates is slightly larger in TimingCamouflage+. 
The comparison of inserted delay units
are shown in \figname~\ref{fig:buffer_DATE}, where the numbers with TimingCamouflage+ are 
consistently smaller in all test cases than with \cite{ZLYPS18}. 
The comparison of area overhead to construct wave-pipelining paths are shown in \figname~\ref{fig:area_DATE}, 
where the area overhead with TimingCamouflage+ is smaller than that with TimingCamouflage in \cite{ZLYPS18}, 
except for s38417. For s1238, s4863, s15850 and s35932, wave-pipelining paths cannot be constructed successfully with 
TimingCamouflage with the maximum numbers of inserted delay units to be no larger than three. 
The power consumption is roughly proportional to the area overhead.

In TimingCamouflage+, the construction of wave-pipelining paths is accelerated by sorting and filtering 
flip-flops described in Section~\ref{sec:workflow}. Consequently, the runtime is 
reduced significantly. The comparison of computation time with TimingCamouflage \cite{ZLYPS18} 
is shown in \figname~\ref{fig:runtime_DATE}. From this comparison, 
it is clear that TimingCamouflage+ outperforms TimingCamouflage in \cite{ZLYPS18} 
in efficiency.%

\section{Conclusion}\label{sec:conclusion}
In this paper, we have proposed TimingCamouflage+ %
to secure circuit netlists against counterfeiting. Since a netlist itself does not
carry all design information anymore, the difficulty of attack has 
increased significantly due to additional test cost and the introduced
wave-pipelining false paths. 
TimingCamouflage+ opens up a new dimension of netlist camouflage at
circuit level, and it
is fully compatible with other previous counterfeiting methods so that 
they can be combined together %
to strengthen netlist security. 
\reviewhl{Future work includes experimental attempts of the attacks discussed in
  Section~\ref{sec:counter_measures} and the  
  combination of the proposed camouflage method with other existing
methods. Advanced timing concepts can be embedded into the netlist to enhance circuit security \cite{ZhangLS16,7105878,Grace18_test,6353349,ZhangLS16_iccad}. 
}


\bibliographystyle{IEEEtran}

\begin{thebibliography}{10}
\providecommand{\url}[1]{#1}
\csname url@samestyle\endcsname
\providecommand{\newblock}{\relax}
\providecommand{\bibinfo}[2]{#2}
\providecommand{\BIBentrySTDinterwordspacing}{\spaceskip=0pt\relax}
\providecommand{\BIBentryALTinterwordstretchfactor}{4}
\providecommand{\BIBentryALTinterwordspacing}{\spaceskip=\fontdimen2\font plus
\BIBentryALTinterwordstretchfactor\fontdimen3\font minus
  \fontdimen4\font\relax}
\providecommand{\BIBforeignlanguage}[2]{{%
\expandafter\ifx\csname l@#1\endcsname\relax
\typeout{** WARNING: IEEEtran.bst: No hyphenation pattern has been}%
\typeout{** loaded for the language `#1'. Using the pattern for}%
\typeout{** the default language instead.}%
\else
\language=\csname l@#1\endcsname
\fi
#2}}
\providecommand{\BIBdecl}{\relax}
\BIBdecl

\bibitem{ZLYPS18}
G.~L. Zhang, B.~Li, B.~Yu, D.~Z. Pan, and U.~Schlichtmann,
  ``Timing{C}amouflage: Improving circuit security against counterfeiting by
  unconventional timing,'' in \emph{Proc. Design, Autom., and Test Europe
  Conf.}, 2018, pp. 91--96.

\bibitem{Arunk17}
A.~Vijayakumar, V.~C. Patil, D.~E. Holcomb, C.~Paar, and S.~Kundu, ``Physical
  design obfuscation of hardware: A comprehensive investigation of device and
  logic-level techniques,'' \emph{{IEEE} Trans. Comput.-Aided Design Integr.
  Circuits Syst.}, vol.~12, no.~1, pp. 64--77, 2017.

\bibitem{BeckerRPB14}
G.~T. Becker, F.~Regazzoni, C.~Paar, and W.~P. Burleson, ``Stealthy
  dopant-level hardware trojans: extended version,'' \emph{J. Cryptographic
  Engineering}, vol.~4, no.~1, pp. 19--31, 2014.

\bibitem{Lap07}
L.-W. Chow, J.~P. Baukus, and J.~William M.~Clark, ``Integrated circuits
  protected against reverse engineering and method for fabricating the same
  using an apparent metal contact line terminating on field oxide,'' in
  \emph{{US Patent} 7,294,935}, 2007.

\bibitem{MalikBPB15}
S.~Malik, G.~T. Becker, C.~Paar, and W.~P. Burleson, ``Development of a
  layout-level hardware obfuscation tool,'' in \emph{Comput. Society Ann. Symp.
  on {VLSI}}, 2015, pp. 204--209.

\bibitem{Rajendran2013}
J.~Rajendran, M.~Sam, O.~Sinanoglu, and R.~Karri, ``Security analysis of
  integrated circuit camouflaging,'' in \emph{Proc. Conf. on Comput. \& Commun.
  Security}, 2013, pp. 709--720.

\bibitem{RajendranSK13}
J.~Rajendran, O.~Sinanoglu, and R.~Karri, ``{VLSI} testing based security
  metric for {IC} camouflaging,'' in \emph{Proc. Int. Test Conf.}, 2013, pp.
  1--4.

\bibitem{LeeT15}
Y.~Lee and N.~A. Touba, ``Improving logic obfuscation via logic cone
  analysis,'' in \emph{Latin-American Test Symp.}, 2015, pp. 1--6.

\bibitem{LiSMZYJP16}
M.~Li, K.~Shamsi, T.~Meade, Z.~Zhao, B.~Yu, Y.~Jin, and D.~Z. Pan, ``Provably
  secure camouflaging strategy for {IC} protection,'' in \emph{Proc. Int. Conf.
  Comput.-Aided Des.}, 2016, pp. 28--35.

\bibitem{Muham16}
M.~Yasin, B.~Mazumdar, O.~Sinanoglu, and J.~Rajendran, ``Camo{P}erturb: Secure
  {IC} camouflaging for minterm protection,'' in \emph{Proc. Int. Conf.
  Comput.-Aided Des.}, 2016, pp. 29:1--29:8.

\bibitem{KavehP17}
K.~Shamsi, M.~Li, T.~Meade, Z.~Zhao, D.~Z. Pan, and Y.~Jin, ``Cyclic
  obfuscation for creating sat-unresolvable circuits,'' in \emph{Proceedings of
  the on Great Lakes Symposium on VLSI}, 2017, pp. 173--178.

\bibitem{Kaveh2017}
------, ``Appsat: Approximately deobfuscating integrated circuits,'' in
  \emph{IEEE International Symposium on Hardware Oriented Security and Trust
  (HOST)}, 2017, pp. 95--100.

\bibitem{Yuanqi2017}
Y.~Shen and H.~Zhou, ``Double {DIP}: Re-evaluating security of logic encryption
  algorithms,'' in \emph{Proceedings of the Great Lakes Symposium on VLSI},
  2017, pp. 179--184.

\bibitem{Abhi16}
A.~Chakraborty, Y.~Liu, and A.~Srivastava, ``Timing{SAT}: Timing profile
  embedded {SAT} attack,'' in \emph{Proc. Int. Conf. Comput.-Aided Des.}, 2018,
  pp. 1--6.

\bibitem{Rajendran2012}
J.~Rajendran, Y.~Pino, O.~Sinanoglu, and R.~Karri, ``Security analysis of logic
  obfuscation,'' in \emph{Proc. Design Autom. Conf.}, 2012, pp. 83--89.

\bibitem{RoyKM08}
J.~A. Roy, F.~Koushanfar, and I.~L. Markov, ``{EPIC:} ending piracy of
  integrated circuits,'' in \emph{Proc. Design, Autom., and Test Europe Conf.},
  2008, pp. 1069--1074.

\bibitem{DupuisBNFR14}
S.~Dupuis, P.~Ba, G.~D. Natale, M.~Flottes, and B.~Rouzeyre, ``A novel hardware
  logic encryption technique for thwarting illegal overproduction and hardware
  trojans,'' in \emph{Int. On-Line Testing Symp.}, 2014, pp. 49--54.

\bibitem{PlazaM15}
S.~M. Plaza and I.~L. Markov, ``Solving the third-shift problem in {IC} piracy
  with test-aware logic locking,'' \emph{{IEEE} Trans. Comput.-Aided Design
  Integr. Circuits Syst.}, vol.~34, no.~6, pp. 961--971, 2015.

\bibitem{Alex2010}
A.~Baumgarten, A.~Tyagi, and J.~Zambreno, ``Preventing {IC} piracy using
  reconfigurable logic barriers,'' \emph{IEEE Design \& Test of Computers},
  vol.~27, pp. 66--75, Feb. 2010.

\bibitem{YA17}
Y.~Xie and A.~Srivastava, ``Delay locking: Security enhancement of logic
  locking against ic counterfeiting and overproduction,'' in \emph{Proc. Design
  Autom. Conf.}, 2017, pp. 1--6.

\bibitem{ChakrabortyB09}
R.~S. Chakraborty and S.~Bhunia, ``{HARPOON:} an obfuscation-based {SoC} design
  methodology for hardware protection,'' \emph{{IEEE} Trans. Comput.-Aided
  Design Integr. Circuits Syst.}, vol.~28, no.~10, pp. 1493--1502, 2009.

\bibitem{Monir2018}
M.~Zaman, A.~Sengupta, D.~Liu, O.~Sinanoglu, Y.~Makris, and J.~J.~V. Rajendran,
  ``Towards provably-secure performance locking,'' in \emph{Proc. Design,
  Autom., and Test Europe Conf.}, 2018, pp. 1558--1101.

\bibitem{GPU18}
A.~Chakraborty, Y.~Xie, and A.~Srivastava, ``{GPU} obfuscation: Attack and
  defense strategies,'' in \emph{Proc. Design Autom. Conf.}, 2018, pp. 1--6.

\bibitem{Nith2018}
N.~G. Jayasankaran, A.~S. Borbon, A.~S. Borbon, J.~Hu, and jeyaviyan Rajendran,
  ``Towards provably-secure analog and mixed-signal locking against
  overproduction,'' in \emph{Proc. Int. Conf. Comput.-Aided Des.}, 2018, pp.
  1--8.

\bibitem{Pra2015}
P.~Subramanyan, S.~Ray, and S.~Malik, ``Evaluating the security of logic
  encryption algorithms,'' in \emph{IEEE International Symposium on Hardware
  Oriented Security and Trust (HOST)}, 2015, pp. 137--143.

\bibitem{MMM2017}
M.~Yasin, B.~Mazumdar, O.~Sinanoglu, and J.~J.~V. Rajendran, ``Security
  analysis of {A}nti-{SAT},'' in \emph{Proc. Asia and South Pacific Des. Autom.
  Conf.}, 2017, pp. 342--347.

\bibitem{Xiaolin2017}
X.~Xu, B.~Shakya, M.~M. Tehranipoor, and D.~Forte, ``Security analysis of
  {A}nti-{SAT},'' in \emph{nternational Conference on Cryptographic Hardware
  and Embedded Systems}, 2017, pp. 1--21.

\bibitem{Dkiro03}
D.~Kirovski and M.~Potkonjak, ``Local watermarks: methodology and application
  to behavioral synthesis,'' \emph{{IEEE} Trans. Comput.-Aided Design Integr.
  Circuits Syst.}, vol.~22, pp. 1277--1283, Sep. 2003.

\bibitem{Fari2001}
F.~Koushanfar and G.~Qu, ``Hardware metering,'' in \emph{Proc. Design Autom.
  Conf.}, 2001, pp. 1--6.

\bibitem{xiaoxiao2017}
X.~Wang, Y.~Guo, T.~Ramhan, D.~Zhang, and M.~Tehranipoor, ``{DOST}: Dynamically
  obfuscated wrapper for split test against {IC} piracy,'' in \emph{IEEE Asian
  Hardware-Oriented Security and Trust Symposium (AsianHOST)}, 2017, pp. 1--6.

\bibitem{Jeya2013}
J.~Rajendran, O.~Sinanoglu, and R.~Karri, ``Is split manufacturing secure?'' in
  \emph{Proc. Design, Autom., and Test Europe Conf.}, 2013, pp. 1259--1264.

\bibitem{Hsieh1995}
H.-Y. Hsieh, W.~Liu, R.~K. Cavin, III, and C.~T. Gray, ``Concurrent timing
  optimization of latch-based digital systems,'' in \emph{Proc. Int. Conf.
  Comput. Des.}, 1995, pp. 680--685.

\bibitem{Burleson1998}
W.~P. Burleson, M.~Ciesielski, F.~Klass, and W.~Liu, ``Wave-pipelining: A
  tutorial and research survey,'' \emph{{IEEE} Trans. {VLSI} Syst.}, vol.~6,
  no.~3, pp. 464--474, Sep. 1998.

\bibitem{SeetharamanV09}
G.~Seetharaman and B.~Venkataramani, ``Automation schemes for {FPGA}
  implementation of wave-pipelined circuits,'' \emph{{ACM} Trans. Reconf. Tech.
  Sys.}, vol.~2, no.~2, 2009.

\bibitem{Grace2018_DAC}
G.~L. Zhang, B.~Li, M.~Hashimoto, and U.~Schlichtmann, ``Virtual{S}ync: Timing
  optimization by sychronizing logic waves with sequential and combinational
  components as delay units,'' in \emph{Proc. Design Autom. Conf.}, 2018, pp.
  1--6.

\bibitem{Yuan2010}
F.~Yuan and Q.~Xu, ``On timing-independent false path identification,'' in
  \emph{Proc. Int. Conf. Comput.-Aided Des.}, 2010, pp. 532--535.

\bibitem{DuYG89}
D.~H. Du, S.~H. Yen, and S.~Ghanta, ``On the general false path problem in
  timing analysis,'' in \emph{Proc. Design Autom. Conf.}, 1989, pp. 555--560.

\bibitem{Heragu1997}
K.~Heragu, J.~H. Patel, and V.~D. Agrawal, ``Fast identification of untestable
  delay faults using implications,'' in \emph{Proc. Int. Conf. Comput.-Aided
  Des.}, 1997, pp. 642--647.

\bibitem{Multivt}
\BIBentryALTinterwordspacing
{Xiaodong Zhang}, ``High performance low leakage design using power compiler
  and {M}ulti-{V}t libraries,'' 2003. [Online]. Available:
  \url{www.synopsys.com}
\BIBentrySTDinterwordspacing

\bibitem{Ruchir2004}
R.~Puri, ``Minimizing power under performance constraint,'' in
  \emph{International Conference on Integrated Circuit Design and technology},
  2004, pp. 159--163.

\bibitem{Frank2005}
F.~G. Frank~Sill and D.~Timmermann, ``Reducing leakage with mixed-vth
  ({MVT}),'' in \emph{International Conference on VLSI Design}, 2005, pp.
  874--877.

\bibitem{Ithi2016}
I.~R. Nirmala, D.~Vontela, S.~Ghosh, and A.~Iyengar, ``A novel threshold
  voltage defined switch for circuit camouflaging,'' in \emph{IEEE European
  Test Symposium (ETS)}, 2016, pp. 1558--1780.

\bibitem{Meng2018}
M.~Li, K.~Shamsi, Y.~Jin, and D.~Z. Pan, ``Timing{SAT}: Decamouflaging
  timing-based logic obfuscation,'' in \emph{Proc. Int. Test Conf.}, 2018, pp.
  1--10.

\bibitem{Mathies2019}
M.~D. Silva, M.-L. Flottes, G.~D. Natale, and B.~Rouzeyre, ``Preventing scan
  attacks on secure circuits through scan chain encryption,'' \emph{{IEEE}
  Trans. Comput.-Aided Design Integr. Circuits Syst.}, vol.~38, no.~3, pp.
  538--550, 2019.

\bibitem{Byang2006}
B.~Yang, K.~Wu, and R.~Karri, ``Secure scan: A design-for-test architecture for
  crypto chips,'' \emph{{IEEE} Trans. Comput.-Aided Design Integr. Circuits
  Syst.}, vol.~25, no.~10, pp. 2287--2293, 2006.

\bibitem{Mohamed2017}
M.~E. Massad, M.~E. Massad, and M.~E. Massad, ``Reverse engineering camouflaged
  sequential circuits without scan access,'' in \emph{Proc. Int. Conf.
  Comput.-Aided Des.}, 2017, pp. 22--40.

\bibitem{Kaveh2019}
K.~Shamsi, M.~Li, D.~Z. Pan, and Y.~Jin, ``{KC2}: Key-condition crunching for
  fast sequential circuit deobfuscation,'' in \emph{Proc. Design, Autom., and
  Test Europe Conf.}, 2019, pp. 534--539.

\bibitem{antisat2019}
Y.~Xie and A.~Srivastava, ``{A}nti-{SAT}: Mitigating {SAT} attack on logic
  locking,'' \emph{{IEEE} Trans. Comput.-Aided Design Integr. Circuits Syst.},
  vol.~38, pp. 199--207, 2019.

\bibitem{Mitigat2016}
------, ``Mitigating {SAT} attack on logic locking,'' in \emph{CHES}, 2016, pp.
  127--146.

\bibitem{Azar2019SMTAN}
K.~Z. Azar, H.~M. Kamali, H.~Homayoun, and A.~Sasan, ``{SMT} attack: Next
  generation attack on obfuscated circuits with capabilities and performance
  beyond the {SAT} attacks,'' in \emph{Conference on Cryptographic Hardware and
  Embedded Systems}, 2019.

\bibitem{Devadas2006}
S.~Devadas, K.~Keutzer, and S.~Malik, ``Computation of floating mode delay in
  combinational circuits: Theory and algorithms,'' \emph{{IEEE} Trans.
  Comput.-Aided Design Integr. Circuits Syst.}, vol.~12, pp. 1913--1923, Nov.
  2006.

\bibitem{Coudert2010}
O.~Coudert, ``An efficient algorithm to verify generalized false paths,'' in
  \emph{Proc. Design Autom. Conf.}, 2010, pp. 188--193.

\bibitem{Kim2003}
K.~S. Kim, S.~Mitra, and P.~G. Ryan, ``Delay defect characteristics and testing
  strategies,'' \emph{{IEEE} Des. Test. Comput.}, vol.~20, pp. 8--16, Sep.
  2003.

\bibitem{Char91}
C.~E. Leiserson and J.~B. Saxe, ``Retiming synchronous circuitry,''
  \emph{Journal Algorithmica}, vol.~6, pp. 5--35, 1991.

\bibitem{gurobi}
\BIBentryALTinterwordspacing
{Gurobi Optimization, Inc.}, ``Gurobi optimizer reference manual,'' 2013.
  [Online]. Available: \url{http://www.gurobi.com}
\BIBentrySTDinterwordspacing

\bibitem{chen2011applied}
D.~Chen, R.~Batson, and Y.~Dang, \emph{Applied Integer Programming: Modeling
  and Solution}.\hskip 1em plus 0.5em minus 0.4em\relax Wiley, 2011.

\bibitem{ZhangLS16}
G.~L. Zhang, B.~Li, and U.~Schlichtmann, ``Sampling-based buffer insertion for
  post-silicon yield improvement under process variability,'' in \emph{Proc.
  Design, Autom., and Test Europe Conf.}, 2016, pp. 1457--1460.

\bibitem{7105878}
B.~{Li} and U.~{Schlichtmann}, ``Statistical timing analysis and criticality
  computation for circuits with post-silicon clock tuning elements,''
  \emph{{IEEE} Trans. Comput.-Aided Design Integr. Circuits Syst.}, vol.~34,
  no.~11, pp. 1784--1797, 2015.

\bibitem{Grace18_test}
G.~L. Zhang, B.~Li, Y.~Shi, J.~Hu, and U.~Schlichtmann, ``Effi{T}est2:
  Efficient delay test and prediction for post-silicon clock skew configuration
  under process varaitions,'' \emph{{IEEE} Trans. Comput.-Aided Design Integr.
  Circuits Syst.}, vol.~38, no.~4, pp. 705--718, 2019.

\bibitem{6353349}
N.~{Chen}, B.~{Li}, and U.~{Schlichtmann}, ``Iterative timing analysis based on
  nonlinear and interdependent flipflop modelling,'' \emph{IET Circuits,
  Devices Systems}, vol.~6, no.~5, pp. 330--337, 2012.

\bibitem{ZhangLS16_iccad}
G.~L. Zhang, B.~Li, and U.~Schlichtmann, ``Piece{T}imer: {A} holistic timing
  analysis framework considering setup/hold time interdependency using a
  piecewise model,'' in \emph{Proc. Int. Conf. Comput.-Aided Des.}, 2016, pp.
  100:1--100:8.

\end{thebibliography}

\begin{footnotesize}
\begin{IEEEbiographynophoto}
{Grace Li Zhang} 
received the Dr.-Ing. from \tum{}, Munich, Germany, in 2018. 
She is currently a postdoctoral researcher with the Chair of Electronic Design Automation, TUM. 
Her current research interests include high-performance and low-power design, as well as emerging systems. 
\end{IEEEbiographynophoto}

\begin{IEEEbiographynophoto}
{Bing Li} 
received the Dr.-Ing. degree from \tum{}, Munich, Germany, in  
2010 and finished the Habilitation there in 2018. He  
is currently a researcher with the Chair of Electronic  
Design Automation, TUM. His current research 
interests include high-performance and low-power  
design of integrated circuits and systems. %

\end{IEEEbiographynophoto}

\begin{IEEEbiographynophoto}
{Meng Li} (S'15--M'18) 
received the Ph.D. degree from the University of Texas at Austin, Austin, Tx, 
USA under the supervision of Dr. D. Z. Pan in 2018. 
He is currently an AI research scientist in Facebook. 
His research interests include AI Software/Hardware Co-design, 
hardware-oriented security and reliability. 
Dr. Li was the recipient of the UT Austin Margarida Jacome Outstanding Dissertation Prize in 2019, 
the EDAA Outstanding Dissertation Award in 2019, 
the First Place in the Grand Final of ACM Student Research Competition in 2018, 
the Best Poster Award in ASPDAC Ph.D. forum in 2018, 
the UT Austin University Graduate Fellowship in 2013, as well as the Best Paper Award in HOST'2017 and GLSVLSI'2018.
\end{IEEEbiographynophoto}

\begin{IEEEbiographynophoto}
{Bei Yu} 
is currently an Assistant Professor in the Department of 
Computer Science and Engineering, The Chinese University of Hong Kong. 
His research interests include machine learning and deep learning with applications in VLSI CAD. 
\end{IEEEbiographynophoto}

\begin{IEEEbiographynophoto}
{David Z. Pan} (S'97--M'00--SM'06--F'14)
is currently Engineering Foundation Professor at the Department of Electrical and Computer Engineering, 
University of Texas at Austin. 
His research interests include cross-layer nanometer IC design for manufacturability,  
reliability, security, machine learning and hardware acceleration, 
design/CAD for analog/mixed signal designs and emerging technologies. 
He has published over 360 journal articles and refereed conference papers. 
He has served in many ACM/IEEE editorial boards and conference committees including various leadership roles. 
He has received numerous awards for his research contributions, 
including SRC Technical Excellence Award, 18 Best Paper Awards, NSF CAREER Award, IBM Faculty Award four times, etc. He is a Fellow SPIE.
\end{IEEEbiographynophoto}

\begin{IEEEbiographynophoto}
{Michaela Brunner}
is currently working as a Ph.D student at the Technical 
University of Munich at the Chair of Security in Information Technology. 
She received the bachelor and master degree in electrical and computer 
engineering from the Technical University of Munich in 2016 and 2018. 
Her primary research interests are in the area of hardware security, 
sequential netlist reverse engineering, including gate-level netlist 
analysis as well as netlist obfuscation, netlist encryption and locking 
methods. She received the VDE award 2018 from the organization of German 
electrical engineers in south Bavaria for her master thesis.
\end{IEEEbiographynophoto}

\begin{IEEEbiographynophoto}
{Georg Sigl}
finished his PhD in Electrical Engineering at Technical University Munich in 1992. 
Afterwards he held several positions in research and development at Siemens and Infineon. 
From 2000 until 2010 he was responsible for the development 
of new secure microcontroller platforms in the Chip Card and Security division. 
Under his responsibility, two award winning platforms have been designed. 
In June 2010, he founded a new chair at Technical University of Munich for Security in Electrical Engineering and Information Technology. 
In parallel, he is driving embedded security research as director at the Fraunhofer Research Institute for Applied and Integrated Security AISEC Munich.
\end{IEEEbiographynophoto}

\begin{IEEEbiographynophoto}
{Ulf Schlichtmann} (S'88--M'90--SM'18) 
received the Dipl.-Ing. and Dr.-Ing. degrees 
in electrical engineering and information technology from Technical University of Munich (TUM), 
Munich, Germany, in 1990 and 1995, respectively. 
He is Professor and the Head of the Chair of Electronic Design Automation at TUM. 
He joined TUM in 2003, following 10 years in industry. 
His current research interests include computer-aided design of electronic circuits and systems, 
with an emphasis on designing reliable and robust systems. 
Recently, he also addresses emerging technologies such as lab-on-chip and photonics.
\end{IEEEbiographynophoto}
\end{footnotesize}
\vfill

\clearpage

\twocolumn[\section*{Wave-pipelining Construction for TimingCamouflage+ \\ 
(Appendix to [R1]) }]

\textbf{Case 1:} The net from gate $g_i$ to gate $g_j$ has the retimed weight $w_r(e_{g_i,g_j})=w(e_{g_i,g_j})+r(g_j)-r(g_i)= 1$,
and thus a retimed flip-flop $\mathit{ff_{k_1}} $exists along this net. The retimed flip-flop is not removed,  %
denoted as $y_{e_{g_i,g_j}}=0$.

In this case, the circuit can operate with a given clock period $T$, if setup and hold time
constraints can be satisfied, expressed as follows
\begin{align} 
      \overline{s}_{g_i}+t_{su}\le T \label{eq:setup}\\
      \underline{s}_{g_i}\ge t_{h} \label{eq:hold}
\end{align}
where (\ref{eq:setup}) represents that %
the latest arrival time at $g_i$ should be stable $t_{su}$ before a rising clock edge.
The earliest arrival time $\overline{s}_{g_i}$ should be larger than $t_{h}$ as shown in (\ref{eq:hold}),
so that the data can be latched by the flip-flop $\mathit{ff_{k_1}}$ reliably.

The latest and the earliest arrival times at gate $g_j$
after the retimed flip-flop is passed through
are expressed as follows
\begin{align} 
      \overline{s}_{g_j}\ge t_{cq}+\xi_{g_j}+d_{g_j} \label{eq:cq_max}\\
      \underline{s}_{g_j} \le t_{cq}+\xi_{g_j}+d_{g_j} \label{eq:cq_min}
\end{align}
where $t_{cq}$ is the clock-to-q delay of the flip-flop.
$\xi_{g_j}$ is introduced to enlarge the delay of wave-pipelining paths, which
can be
implemented by lengthening interconnects.
$d_{g_j}$
represents the
delay from the input pin to the output pin of $g_j$, which
is set individually for each input of $g_j$.

To guarantee that data traveling from $\mathit{ff_{k_1}}$ and $\mathit{ff_{k_2}}$ are latched by $\mathit{ff_{k_3}}$
correctly,
the latest and the earliest arrival times at $g_j$ should satisfy the following constraints
\begin{align} 
      \overline{s}_{g_j}+t_{su}\le T \label{eq:setup_j}\\
      \underline{s}_{g_j}\ge t_{h} \label{eq:hold_j}
\end{align}
where $\overline{s}_{g_j}$ and $\underline{s}_{g_j}$ can be
replaced by (\ref{eq:cq_max})--(\ref{eq:cq_min}), so that (\ref{eq:setup_j})--(\ref{eq:hold_j})
can be converted as follows
\begin{align} 
      t_{cq}+\xi_{g_j}+d_{g_j}+t_{su}\le T \label{eq:setup_co}\\
      t_{cq}+\xi_{g_j}+d_{g_j}\ge t_{h} \label{eq:hold_co}.
\end{align}

\textbf{Case 2:} %
The net from gate $g_i$ to $g_j$ has the retimed weight $w_r(e_{g_i,g_j})=w(e_{g_i,g_j})+r(g_j)-r(g_i)=1$,
but the flip-flop is removed, denoted as $y_{e_{g_i,g_j}}=1$.  %

In this case, wave-pipelining paths can be constructed between the fanin flip-flops and $\mathit{ff_{k_3}}$,
as in \figname~\ref{fig:shift_arrival}.
After $\mathit{ff_{k_1}}$ is removed,
the data coming from the fanin flip-flops pass the removal point directly and
travel through the gate $g_j$ afterwards.  %
The data should arrive at $\mathit{ff_{k_3}}$ after the first rising clock edge and
before the second rising clock edge to guarantee it is latched correctly [R2].
With respect to wave-pipelining,
the arrival times at $g_j$ %
should satisfy the following constraints %

\begin{align} \label{eq:short_path_new}
\overline{s}_{g_j}+t_{su} \le 2T\\
\label{eq:long_path_new}
\underline{s}_{g_j}-t_h \ge T.
\end{align}

Since
the retimed flip-flop on the net between $g_i$ and $g_j$ is removed,
the data at the output of $g_i$ travels through $g_j$ directly.
Therefore,
the relations between their arrival times
are established as follows
\begin{align} 
      \overline{s}_{g_j}\ge \overline{s}_{g_i}+\xi_{g_j}+d_{g_j} \label{eq:n_max}\\
      \underline{s}_{g_j} \le \underline{s}_{g_i}+\xi_{g_j}+d_{g_j}.\label{eq:n_min}
\end{align}

By replacing $\overline{s}_{g_j}$ and $\underline{s}_{g_j}$
by (\ref{eq:n_max})--(\ref{eq:n_min}), (\ref{eq:short_path_new})--(\ref{eq:long_path_new})
can be converted as follows
\begin{align} 
      \overline{s}_{g_i}+\xi_{g_j}+d_{g_j}+t_{su} \le 2T \label{eq:r_max}\\
      \underline{s}_{g_i}+\xi_{g_j}+d_{g_j}-t_h \ge T\label{eq:r_min}
\end{align}
and thus
\begin{align} \label{eq:c_short_path}
\overline{s}_{g_i}+\xi_{g_j}+d_{g_j}-T+t_{su} \le T\\
\label{eq:c_long_path}
\underline{s}_{g_i}+\xi_{g_j}+d_{g_j}-T \ge t_h 
\end{align}
where $\overline{s}_{g_i}+\xi_{g_j}+d_{g_j}-T$ and $\underline{s}_{g_i}+\xi_{g_j}+d_{g_j}-T$
indicate that the removal of a flip-flop leads
to the shift of the arrival times by a clock
period $T$ after the removal point is passed through.

Besides wave-pipelining paths arriving at $\mathit{ff_{k_3}}$,
single-period paths exist between $\mathit{ff_{k_2}}$ and $\mathit{ff_{k_3}}$.
These paths require that the latest and the earliest arrival times at $g_j$ satisfy the constraints
(\ref{eq:setup_j})--(\ref{eq:hold_j}).
To guarantee that the data from wave-pipelining and single-period paths
are latched correctly at $\mathit{ff_{k_3}}$,
the latest arrival times at $g_j$, $\max\{\overline{s}_{g_i}+\xi_{g_j}+d_{g_j}-T, t_{cq}+\xi_{g_j}+d_{g_j}\}$,
and the earliest arrival times at $g_j$, $\min\{\underline{s}_{g_i}+\xi_{g_j}+d_{g_j}-T,t_{cq}+\xi_{g_j}+d_{g_j}\}$,
should satisfy the following constraints

\begin{align} \label{eq:f_short_path}
\max\{\overline{s}_{g_i}+\xi_{g_j}+d_{g_j}-T, t_{cq}+\xi_{g_j}+d_{g_j}\}+t_{su} \le T\\
\label{eq:f_long_path}
\min\{\underline{s}_{g_i}+\xi_{g_j}+d_{g_j}-T,t_{cq}+\xi_{g_j}+d_{g_j}\} \ge t_h.
\end{align}

\textbf{Case 3:}
The net from gate $g_i$ to $g_j$ does not have a retimed flip-flop, $w_r(e_{g_i,g_j})= 0$. In
this case, the data at the output of $g_i$ passes through $g_j$ directly.
Therefore,
the relation between the arrival times of gate $g_i$ and $g_j$ can be
established as  (\ref{eq:n_max})--(\ref{eq:n_min}).

The wave-pipelining constraints in (\ref{eq:short_path})--(\ref{eq:long_path}) for all constructed paths
are
met by guaranteeing the longest and the shortest constructed paths as wave-pipelining
with (\ref{eq:f_short_path})--(\ref{eq:f_long_path}).
In addition,
all constructed wave-pipelining paths should also meet the gray region (3),
so that attackers cannot determine whether they are wave-pipelining or simple critical paths [R3].
For example, the wave-pipelining paths from fanin flip-flops to $\mathit{ff_{k_3}}$
after $\mathit{ff_{k_1}}$ is removed in \figname~\ref{fig:shift_arrival} should satisfy the
gray region constraint. To achieve this goal,
with respect to wave-pipelining, the
arrival times at $g_j$ should meet the following constraints
\begin{equation}\label{eq:gray_arr}
(1-\tau)\overline{s}_{g_j}\le T\le (1+\tau)\underline{s}_{g_j}
\end{equation}
where the $\overline{s}_{g_j}$ and $\underline{s}_{g_j}$ can
be replaced by those in (\ref{eq:n_max})--(\ref{eq:n_min}),
to convert (\ref{eq:gray_arr}) as follows
\begin{equation}\label{eq:gray_arr_c}
(1-\tau)(\overline{s}_{g_i}+\xi_{g_j}+d_{g_j})\le T \le (1+\tau)(\underline{s}_{g_i}+\xi_{g_j}+d_{g_j}).
\end{equation}

\vspace{1cm}

\noindent References
\vspace{0.3cm}

\footnotesize{
  
 \noindent  [R1] G. L. Zhang, B. Li, M. Li, B. Yu, D. Z. Pan, M. Brunner, G. Sigl and U. Schlichtmann, ``TimingCamouflage+: Netlist Security Enhancement with Unconventional Timing'', DOI:10.1109/TCAD.2020.2974338
 \vskip 3pt 
\noindent [R2] G. L. Zhang, B. Li, M. Hashimoto and U. Schlichtmann, ``VirtualSync: Timing Optimization by Synchronizing Logic Waves with Sequential and Combinational Components as Delay Units'', ACM/IEEE Design Automation Conference (DAC), 2018
 \vskip 3pt 
\noindent [R3] G. L. Zhang, B. Li, B. Yu, D. Z. Pan and U. Schlichtmann, ``TimingCamouflage: Improving Circuit Security against Counterfeiting by Unconventional Timing'', Design, Automation and Test in Europe (DATE), 2018, pp. 91--96}

\end{document}